\documentclass[12pt]{article}

\usepackage{latexsym,amssymb,epsfig}
\input{diagrams}

\DeclareFontFamily{OT1}{rsfs10}{}
\DeclareFontShape{OT1}{rsfs10}{m}{n}{ <-> rsfs10 }{}
\DeclareMathAlphabet{\mathscript}{OT1}{rsfs10}{m}{n}

\newcommand{\eref}[1]{(\ref{#1})}
\newcommand{\sref}[1]{Section~\ref{#1}}
\newcommand{\tref}[1]{Table~\ref{#1}}
\newcommand{\fref}[1]{Figure~\ref{#1}}
\newcommand{\cref}[1]{Chapter~\ref{#1}}
\newcommand{\bcenter}{\begin{center}}
\newcommand{\ecenter}{\end{center}}
\newcommand{\beq}{\begin{equation}}
\newcommand{\eeq}{\end{equation}}
\newcommand{\bea}{\begin{eqnarray}}
\newcommand{\eea}{\end{eqnarray}}
\newcommand{\bean}{\begin{eqnarray*}}
\newcommand{\eean}{\end{eqnarray*}}
\newcommand{\ba}{\begin{array}}
\newcommand{\ea}{\end{array}}
\newcommand{\ben}{\begin{enumerate}}
\newcommand{\een}{\end{enumerate}}
\newcommand{\bi}{\begin{itemize}}
\newcommand{\ei}{\end{itemize}}
\newcommand{\bd}{\begin{description}}
\newcommand{\ed}{\end{description}}
\def\fnote#1#2{\begingroup\def\thefootnote{#1}\footnote{#2}
     \addtocounter{footnote}{-1}\endgroup}

\def\IC{\mathbb{C}}

\def\IR{\mathbb{R}}

\def\IZ{\mathbb{Z}}

\def\IP{\mathbb{P}}

\def\Hom{{\rm Hom}}
\def\dim{{\rm dim}}

\def\cM{{\mathcal M}}

\def\cS{{\mathcal S}}
\def\cO{{\mathcal O}}

\def\cF{{\mathcal F}}
\def\cP{{\mathcal P}}

\def\cV{{\mathcal V}}
\def\nn{\nonumber}
\def\td{\mbox{td}}
\def\ch{\mbox{ch}}
\def\rk{\mbox{rk}}
\def\ind{\mbox{ind}}
\def\dirac{\slash{\! \! \! \! D}}

\def\ad{{\mathop {\rm ad}}}

\def\av{\wedge^2 V}
\def\avt{\wedge^2 \tilde{V}}

\def\to{\rightarrow}
\def\b{\beta}

\newcommand{\mat}[1]{\left( \matrix{#1} \right)}

\def\II{{\rlap{1} \hskip 1.6pt \hbox{1}}}

\newarrow{DDTo}{}{dash}{}{dash}>
\def\ol{\overline}
\def\tx{\tilde{X}}
\def\tv{\tilde{V}}
\def\ext{{\rm Ext}}

\def\op1{{\mathcal O}_{\IP^1}}
\def\opt1{{\mathcal O}_{\IP^1_t}}
\def\opx1{{\mathcal O}_{\IP^1_x}}
\def\obp{{\mathcal O}_{B'}}
\newcommand{\uptimes}[1]{\stackrel{#1}{\times}}
\def\v23{V_2 \otimes V_3}

\def\pt{{\rm pt}}
\def\GRR{{Groethendieck-Riemann-Roch}}
\def\q{{\sf Q} }
\def\sm{(SU(3)_C \times SU(2)_L \times U(1)_Y)/\IZ_6}

\newcommand\href[2]{#2}

\begin{document}

\begin{titlepage}

\vspace{-2cm}

\title{
   \hfill{\normalsize  UPR-1097-T} \\[1em]
   {\LARGE The Spectra of Heterotic Standard Model Vacua
\author{Ron Donagi$^1$, Yang-Hui He$^2$, Burt A.~Ovrut$^2$, 
        and Ren\'{e} Reinbacher$^3$
        \fnote{~}{donagi@math.upenn.edu;
        yanghe, ovrut@physics.upenn.edu;
        rreinb@physics.rutgers.edu}\\[0.5cm]
   {\normalsize $^1$ 
        Department of Mathematics, University of Pennsylvania} \\
        {\normalsize Philadelphia, PA 19104--6395, USA} \\
   {\normalsize $^2$
        Department of Physics, University of Pennsylvania} \\
   {\normalsize Philadelphia, PA 19104--6396, USA} \\
   {\normalsize $^3$
        Department of Physics and Astronomy, Rutgers University}\\
   {\normalsize Piscataway, NJ 08855-0849, USA}}
}
\date{}
}

\maketitle

\begin{abstract}
A formalism for determining the massless spectrum of a class of
realistic heterotic string vacua is presented. 
These vacua, which
consist of $SU(5)$ holomorphic bundles on torus-fibered Calabi-Yau
threefolds with fundamental group $\IZ_2$, lead to low energy theories
with standard model gauge group $\sm$
and three families of quarks and leptons.
A methodology for determining the sheaf cohomology of these bundles
and the representation of $\IZ_2$ on each cohomology group is given.
Combining these results with the action of a $\IZ_2$ Wilson line, we
compute, tabulate and discuss the massless spectrum.
\end{abstract}

\thispagestyle{empty}

\end{titlepage}

%%%%%%%%%
%		INTRO
%%%%%%%%%%%%%
\section{Introduction}
The early discussions of realistic vacua in heterotic superstring
theory were within the context of the ``standard embedding''
\cite{GSW}
of the spin connection into the gauge connection. Said differently,
these vacua always involve a holomorphic $E_8$ vector bundle, $V$,
which is induced by the tangent bundle $TX$ of the smooth Calabi-Yau
threefold $X$. %; that is, $V=TX$. 
Although leading to interesting low
energy physics, this approach suffers from the fact that it is highly
constrained, the tangent bundle being only one out of an enormous
number of possible holomorphic bundles $V$. One consequence of this
constraint is the fact that all heterotic vacua based on the standard
embedding require the spontaneous breaking of $E_8$ to $E_6$, which is
then further broken by Wilson lines. Although $E_6$ is a possible
grand unified group, other groups, such as $SU(5)$ or $Spin(10)$, are
simple and more compelling given recent experimental data.
Equally significant is that, in the standard embedding, the low energy
spectrum and couplings are completely determined by the cohomology of
the tangent bundle $TX$. This seriously constrains these quantities,
and it has been difficult to find realistic models in this context.

A technical breakthrough in this regard was presented in
\cite{FMW1,D,FMW2}, 
where it was shown how to construct a large class of
stable, holomorphic vector bundles on simply connected elliptically
fibered Calabi-Yau threefolds where $V \ne TX$. Such bundles admit
connections satisfying the hermitian Yang-Mills equations. This work
was extended in \cite{Donagi:1998xe}-\cite{Donagi:1999gc}, 
and it was shown that these bundles can lead
to heterotic string vacua with a wide range of low energy gauge
groups, including $SU(5)$ and $Spin(10)$. Many of the physical
properties of these vacua have been studied, including supersymmetry
breaking \cite{Lukas:1999kt,Lalak1}, 
the moduli space of the vector bundle
\cite{Buchbinder:2002pr}-\cite{moduli2},  and,
in the strongly coupled case, the associated M5-brane moduli space
\cite{Donagi:1999jp}, 
small instanton phase transitions
\cite{Ovrut:2000qi}-\cite{Douglas:2004yv}, 
non-perturbative superpotentials
\cite{Buchbinder:2002pr,Buchbinder:2002ic,Lima:2001nh,Lima:2001jc}, 
and fluxes \cite{Krause:2000gp}-\cite{Cardoso:2002hd}.
More recently, it was
shown how to compute the sheaf cohomology of $V$ and its tensor
products, thus determining the complete particle physics spectrum
\cite{spec1,spec}. An important conclusion of these papers is that
the spectrum depends on the region of vector bundle moduli space in
which it is evaluated. Although constant for generic moduli, the
spectrum can jump dramatically on subspaces of co-dimension one or
higher always containing, however, three families of quarks and
leptons. These vacua also underlie the theory of brane universes
\cite{Lukas:1997fg}-\cite{Lukas:1998yy} and ekpyrotic and Big Crunch/Big Bang 
cosmology \cite{Khoury:2001zk}-\cite{Khoury:2001wf}. 
The major drawback of these vacua is that the compactification
manifold is simply connected. It follows that these are all GUT
theories which cannot be broken to the standard model with Wilson
lines \cite{Sen:1985eb}-\cite{Andreas}.
Although many of these vacua contain Higgs multiplets whose
vacuum expectation values could induce symmetry breaking,
it would be simpler and more natural if Wilson lines could be introduced.

This was accomplished in \cite{Donagi:2000fw}-\cite{Donagi:1999ez}, 
where stable holomorphic vector
bundles with structure group $SU(5)$ were constructed over
torus-fibered Calabi-Yau threefolds with fundamental group $\pi_1(X) =
\IZ_2$. These heterotic vacua lead, using a $\IZ_2$ Wilson line, to
low energy theories that are anomaly free, have three families of
quarks/leptons and the gauge group $\sm$. 
This work was extended 
to vector bundles with
structure group $SU(4)$ on torus-fibered  Calabi-Yau threefolds with
$\pi_1(X) = \IZ_2 \times \IZ_2$ in
\cite{Donagi:2003tb}-\cite{Ovrut:2002jk} and $\pi_1(X) = \IZ_3 \times
\IZ_3$ in \cite{Braun:2004xv}. Although very promising, it is
essential that one now compute the exact spectrum and couplings in
these standard model vacua. In this paper, we take a major step
in this direction by computing the particle spectrum for the
vacua in \cite{Donagi:2000fw}-\cite{Donagi:1999ez}.

This is accomplished as follows. In
\cite{Donagi:2000fw}-\cite{Donagi:1999ez},  $X$ is the quotient $X =
\tx / \IZ_2$, where $\tx$ is a simply connected Calabi-Yau
threefold. Denote by $\tv$ the pull-back of $V$ to $\tv$. To find the
particle spectrum, one must first compute the sheaf cohomology of
$\tv$ and its tensor products. This is a non-trivial task involving
various techniques in cohomological algebra and algebraic geometry. In
this paper, we present a systematic approach to such computations, and
determine all relevant cohomology groups in our theory. The next step
is to find the explicit representations of $\IZ_2$ in each of these
spaces. We give a precise methodology for accomplishing this. This
approach is then used to determine each of the requisite $\IZ_2$
representations. The above information, in conjunction with the action
of the $\IZ_2$ Wilson line, can be utilized to find all group
multiplets that are invariant under $\IZ_2$, as well as their
multiplicities. When constructing the quotient Calabi-Yau threefold $X
= \tx / \IZ_2$, these invariant multiplets descend to $X$ and form the
massless particle physics spectrum. 
Using these techniques, we compute and tabulate the spectrum.

Specifically, we do the following. In Section 2, we present a general
formalism for describing $G( \subset E_8)$-bundles, Wilson lines and
the massless spectrum associated with non-simply connected Calabi-Yau
threefolds $X$ with $\pi_1(X) = F$. It is shown that determining this
spectrum requires the computation of specific sheaf cohomologies on
the covering Calabi-Yau threefold $\tx$, as well as the action of $F$
on these groups. This formalism is illustrated for several values of
$F$, including $F = \IZ_2$. Section 3 is devoted to a brief review of
the results in \cite{Donagi:2000fw}-\cite{Donagi:1999ez}.
Specifically, we discuss the construction of
torus-fibered Calabi-Yau threefolds $X$ with fundamental group $F =
\IZ_2$. It is shown how to construct stable, holomorphic bundles $V$
with structure group $SU(5)$ on $X$. These arise from $\IZ_2$
invariant bundles $\tv$ on $\tx$ and satisfy the basic
phenomenological constraints of particle physics. Computing the
massless spectrum of this theory requires determining the sheaf
cohomology of $\tv$ and its tensor products. A general method for
doing this is presented in Section 4 and used to compute the relevant
cohomology groups in our theory. Section 5 is devoted to finding the
explicit representations of $\IZ_2$ on these cohomology
groups. Combining the results of Section 5 with the $F = \IZ_2$
example in Section 2, the massless spectrum of our theory is computed,
tabulated and discussed in Section 5. Finally, in the Appendix 
we present some useful mathematical facts used throughout the paper.

%%%%%%%
%
%
%
%%%%%%%
\section{The Spectra of Heterotic Compactifications \\
	with Wilson Lines}\label{s:1}
A vacuum in weakly coupled heterotic string theory is specified by a
pair $(X,\ol{\cV})$, where $X$ is a Calabi-Yau threefold and
$\ol{\cV}$ is a stable $E_8 \times E_8$ holomorphic principal bundle
on $X$ satisfying the Green-Schwarz anomaly cancellation condition
\cite{GS}
\beq
c_2(\ol{\cV}) = c_2(TX) \ .
\eeq 
Note that 
specifying the $E_8 \times E_8$ bundle $\ol{\cV}$ is
the same as giving two $E_8$ bundles $\cV$ and $\cV''$.
The anomaly cancellation condition can be written as
\beq\label{anom1}
c_2(\cV) + c_2(\cV'') = c_2(TX) \ .
\eeq
In  this
work, we will always take $\cV''$ to be trivial.
Then, condition \eref{anom1} becomes
\beq\label{anom}
c_2(\cV) = c_2(TX) \ .
\eeq
In heterotic M-theory compactifications, this condition is relaxed to
\beq\label{anom2}
c_2(\cV) + [C] = c_2(TX) \ ,
\eeq
where $\cV$ is a stable holomorphic $E_8$ principal bundle in the 
observable sector and
$[C]$ is the class of some effective curve $C \subset X$ on which
M5-branes wrap. 

The particle
spectrum of this compactification consists \cite{GSW} of
zero-modes of the ten-dimensional Dirac operator
\beq
\dirac : \Gamma(\ad \cV \otimes S^+_{10}) \to 
	\Gamma(\ad \cV \otimes S^-_{10}) \ .
\eeq
Here $\ad\cV$ is the rank-248 vector bundle associated to $\cV$ by the
adjoint representation of $E_8$, $S^{\pm}_{10}$ are the bundles of
positive and negative chirality spinors in 10-dimensions, and 
$\Gamma$ denotes global sections of a bundle over the 10-dimensional
space $\IR^4 \times X$. 
(Note that we can consider $\ad \cV$ to be a bundle on $\IR^4 \times
X$ by simply pulling it back from $X$).

The 10-dimensional spinors decompose in terms of their (Minkowski)
$\IR^4$ and (internal) $X$ components as
\beq
S^+_{10} = (S^+_4 \otimes S^+_6) \oplus
		(S^-_4 \otimes S^-_6).
\eeq
The internal spinors, on the Calabi-Yau threefold $X$, can be
identified with the $(0,q)$ forms ${\cal A}^{0,q}$ on $X$, with even/odd $q$
corresponding to positive/negative chirality:
\beq
S^+_6 \simeq {\cal A}^{0,0} \oplus {\cal A}^{0,2}, \qquad
S^-_6 \simeq {\cal A}^{0,1} \oplus {\cal A}^{0,3}.
\eeq
In terms of this identification, the Dirac operator becomes 
$\dirac = \ol{\partial} + \ol{\partial}^* + \dirac_4$ 
coupled to $\ad \cV$, where
$\ol{\partial}$ is the Dolbeault operator on $X$, and $\dirac_4$ is
the Dirac operator on flat $\IR^4$. 
Putting these facts together, we find that the spectrum is
\beq
\ker(\dirac) \simeq 
\left( \bigoplus_{q=0,2} H^q(X, \ad \cV) \otimes {\bf S}^+_4
%\Gamma(\IR^4, S^+_4) 
	\right)
\oplus
\left( \bigoplus_{q=1,3} H^q(X, \ad \cV) \otimes {\bf S}^-_4
%\Gamma(\IR^4, S^-_4) 
	\right),
\eeq
where ${\bf S}^{\pm}_4$ denote the constant sections of the bundle
$S^{\pm}_4$ on $\IR^4$.
The positive chirality particles are those which multiply
${\bf S}^+_4$, so they are given by (a basis of)
\beq\label{poschiral}
\bigoplus_{q=0,2} H^q(X, \ad \cV).
\eeq
Their negative chirality anti-particles are similarly given by a basis
of
\beq\label{negchiral}
\bigoplus_{q=1,3} H^q(X, \ad \cV).
\eeq
By Serre duality, this is the dual space to \eref{poschiral}, as it
should be by CPT. 
Recall that, for each charged particle, CPT predicts the existence of an 
anti-particle of opposite charge. 
The annihilation of a particle with its 
anti-particle can be interpreted as a natural pairing. 
Hence, we can 
interpret the space of anti-particles as the dual of the space of 
particles.
In order to describe the complete spectrum, we will
in this work calculate
\beq
Spec := \bigoplus_{q=0,1} H^q(X, \ad \cV).
\eeq
Then, $\ker(\dirac)$ is obtained by adding the duals to $Spec$.
%the cohomologies for $q=0$ (which is dual to
%$q=3$) and for $q=1$ (which is dual to $q=2$).

In practice, the $E_8$ bundle $\cV$ is often associated to some
stable $G$-bundle $V$ on $X$, 
where $G \subset E_8$ is some subgroup, e.g.,
$G = SU(n)$ for $n=3,4$ or $5$~\footnote{
	Since all of our bundles are holomorphic, the relevant
	structure groups are actually $G = SL(n, \IC)$. However, to
	conform to the usual physics notation, we will throughout this
	paper refer to these groups as $G=SU(n)$.
}:
\beq\label{def-cV}
\cV = V \uptimes{G} E_8 \ .
\eeq
The resulting compactification then has a low energy gauge group
\beq\label{GH}
H = Z_{E_8}(G) \ ,
\eeq
the commutant of $G$ in $E_8$. The decomposition of the
248-dimensional representation $\ad E_8$ under the product $G \times
H$ then gives an associated decomposition for $\ad \cV$ and the
Dirac-operator zero-modes. For example, we can take $V$ to be an
$SU(3)$ bundle, or equivalently, a rank 3 vector bundle with trivial
determinant.
The usual embedding of $G = SU(3)$ into $E_8$ has commutant $H=E_6$. The
decompostion of $\ad E_8$ into irreducible representations
of $SU(3) \times E_6$ involves four terms
\beq\label{248-3}
248 = 
(1,78) \oplus (8,1) \oplus (3, 27) \oplus (\ol{3}, \ol{27}) \ .
\eeq
Here, 8 and 78 are the adjoints of $SU(3)$ and $E_6$ respectively, 
3 is the
fundamental representation of $SU(3)$, and 27, $\ol{27}$ are the
smallest representations of $E_6$. For the zero-modes we get:
\beq\label{3decomp}
Spec = 
\left( H^0(X, \cO_X) \otimes 78 \right) \oplus
\left( H^1(X, \ad V) \otimes \II \right) \oplus
\left( H^1(X, V) \otimes 27 \right) \oplus
\left( H^1(X, V^*) \otimes \ol{27} \right) \ .
\eeq
Here we think of $V$ as a rank 3 vector bundle on $X$ associated to
the principal $SU(3)$ bundle by the fundamental representation,
$V^*$ is its dual vector bundle, $\ad V$ is the rank-8 vector
bundle of traceless endomorphisms of $V$, and $\cO_X$ is the trivial
rank-1 bundle on $X$. Note that the stability of $V$ and the
Calabi-Yau property of $X$ guarantee that for each of the associated
bundles $(\cO_X, \ad V, V, V^*)$, the cohomology can be non-zero
for either $q=0$ or $q=1$ but not both, as indicated in
\eref{3decomp}.

As another example, we consider the usual embedding of $G=SU(5)$ into
$E_8$. The commutant is $H=SU(5)$ and 
the $SU(5)_G \times SU(5)_H$-decomposition is 
\beq\label{E8-su5}
248 =
(1,24) \oplus (24,1) 
\oplus (10,5) \oplus (\ol{10},\ol{5})
	\oplus (5, \ol{10}) \oplus (\ol{5},10) \ .
\eeq
The zero-modes are
\bea
Spec &=& 
\left( H^0(X, \cO_X) \otimes 24 \right) \oplus
\left( H^1(X, \ad V) \otimes \II \right) \oplus
\left( H^1(X, \av) \otimes 5 \right) \oplus
\left( H^1(X, \av^*) \otimes \ol{5} \right) \nn \\
&& \oplus \left( H^1(X, V) \otimes \ol{10} \right) \oplus
\left( H^1(X, V^*) \otimes 10 \right) \ .
\eea
More generally, for $G \subset E_8$ with commutant $H$, we write
\beq\label{adE8}
\ad E_8 = \bigoplus_i U_i \otimes R_i \ ,
\eeq
where $U_i$ runs over irreducible representations of $G$, and $R_i$ are
corresponding representations of $H$. 
Using this decomposition of the representation $\ad E_8$ on each fiber
of  the $E_8$ bundle defined in \eref{def-cV}, 
we find the decomposition
\beq\label{decomp}
\ad \cV =  \bigoplus_i U_i(V) \otimes R_i \ ,
\eeq
where $U_i(V)$ are the vector bundles associated to the $G$-bundle
$V$ via the representations $U_i$ of $G$.

Next we want to see how these results are modified by Wilson
lines. Let $F \subset H$ be a finite subgroup which acts on a
Calabi-Yau threefold $\tilde{X}$ freely with a Calabi-Yau quotient
$X = \tilde{X} / F$.
The $G$-bundle $V$ and the $E_8$-bundle
$\cV = V \uptimes{G} E_8$ on $X$ pull back to a $G$-bundle $\tilde{V}
= p^* V$ and an $E_8$-bundle
$\tilde{\cV} = p^*\cV = \tilde{V} \uptimes{G} E_8$
on $\tilde{X}$, where $p : \tilde{X} \to X$ is the covering map. 
The action of $F$ on $\tilde{X}$ lifts to actions, denoted $\rho$,  
on $\tv$, $\tilde{\cV}$, hence on their cohomologies. 
The cohomology group computed on $X$ is precisely the
$\rho(F)$-invariant part of the cohomology on $\tilde{X}$
\beq\label{invcoho}
H^q(X, \ad \cV) = H^q(\tilde{X}, \ad \tilde{\cV})^{\rho(F)} \ .
\eeq

The Wilson line $W$ is the flat $H$-bundle on $X$ induced from the
$F$-cover $p: \tilde{X} \to X$ via the given embedding of $F$ in $H$:
\beq\label{wilson}
W := \tilde{X} \uptimes{F} H \ .
\eeq
The $(G \times H)$-bundle $V \oplus W$ induces another $E_8$-bundle
on $X$:
\beq\label{VpGH}
\cV' = (V \oplus W) \uptimes{(G \times H)} E_8 \ .
\eeq
Our goal in this work is to study the particle spectrum and other
properties of the heterotic vacuum given by compactification on
$(X,\cV')$. Since the structure group of $\cV'$ can be reduced to $G
\times F$ (but not to $G$), we see in analogy with \eref{GH} that this
vacuum has low energy gauge group
\beq\label{S}
S := Z_H(F) = Z_{E_8} (G \times F) \ .
\eeq

We will work primarily with a particular class of geometric examples
which is reviewed in Section 2. In the remainder of the present
section we will describe the general approach. This is based on the
observation that, when pulled backed to $\tilde{X}$, the two bundles
$\cV$, $\cV'$ coincide:
\beq
p^* \cV' \simeq p^* \cV = \tilde{\cV} \ .
\eeq
This is because 
the finite structure group $F$ of the Wilson line $W$ is killed
in the passage from $X$ to $\tilde{X}$. Another way to describe this
is to note that there are two actions $\rho$, $\rho'$ of $F$ on
$\tilde{\cV}$, both lifting the given $F$ action on $\tilde{X}$. 
The quotient by $\rho$ gives $\cV$, and the quotient by $\rho'$ gives
$\cV'$. 
The analogue of \eref{invcoho} is:
\beq\label{invcoho2}
H^q(X, \ad \cV') = H^q(\tilde{X}, \ad \tilde{\cV})^{\rho'(F)} \ .
\eeq
We can write the decomposition \eref{decomp} on $\tilde{X}$:
\beq
\ad \tilde{\cV} =  \bigoplus_i U_i(\tilde{V}) \otimes R_i
\eeq
and use formulas \eref{invcoho}, \eref{invcoho2} to descend to
$X$. The $\rho$ action of $F$ acts only on the associated vector
bundles $U_i(\tilde{V})$, hence on their cohomology, so:
\beq\label{Hdecomp}
H^q(X, \ad \cV) = \bigoplus_i
H^q(\tilde{X}, U_i(\tilde{V}))^{\rho(F)} 
\otimes R_i
\ .
\eeq
The $\rho'$ action of $F$ is a combination of the
$\rho$ action on the $U_i(\tilde{V})$ with the action of $F \subset H$
on the $R_i$:
\beq\label{Hdecomp2}
H^q(X, \ad \cV') = \bigoplus_i
\left( H^q(\tilde{X}, U_i(\tilde{V})) \otimes R_i \right)^{\rho'(F)} \ .
\eeq
Recall that $H^q(X, \ad \cV)$ and its decomposition \eref{Hdecomp}
carry an action of $H$ (which is the natural action on $R_i$ in
\eref{Hdecomp}), but only the subgroup $S \subset H$ acts
on $H^q(X, \ad \cV')$ and its decomposition \eref{Hdecomp2}. To make
the latter more explicit, we decompose each $H$-representation $R_i$
in terms of the irreducible $F$-representations $A_j$:
\beq\label{Ri}
R_i = \bigoplus_j (A_j \otimes B_{ij}), \quad
B_{ij} := \Hom_F(A_j, R_i) \ .
\eeq
Our formula \eref{Hdecomp2} for the particle spectrum then becomes
\beq\label{spec}
H^q(X, \ad \cV') = \bigoplus_{i,j}
(H^q(\tilde{X}, U_i(\tilde{V})) \otimes A_j)^{\rho'(F)} 
	\otimes B_{ij} \ .
\eeq
Here each $B_{ij}$ carries a representation of the low energy gauge
group $S$, which occurs in $H^q(X, \ad \cV')$ with multiplicity
$m_{ij}$ equal to the dimension of the space of $F$-invariants in 
$H^q(\tilde{X}, U_i(\tilde{V})) \otimes A_j$. Note that the
$S$-representation $B_{ij}$ is often not irreducible. Rather, we
should think of $B_{ij}$ as a block of irreducible
$S$-representations, each of which corresponds to some particle. All
the particles in a given block $B_{ij}$ occur in the spectrum with the
same multiplicity $m_{ij}$.

We can summarize our procedure so far as follows. The input involves
\begin{itemize}
\item a structure group $G \subset E_8$,
\item a finite subgroup $F$ of the commutant $H = Z_{E_8}(G)$,
\item a free action of $F$ on a Calabi-Yau threefold $\tilde{X}$ with
	Calabi-Yau quotient $X = \tilde{X} / F$, and
\item a $G$-bundle $V$ on $X$ satisfying the anomaly cancellation
	condition \eref{anom2}.
\end{itemize}

These data determine a Wilson line $W$ on $X$ (as in \eref{wilson}) and
a heterotic vacuum $(X,\cV')$ where $\cV'$ combines the $G$-bundle $V$
with the Wilson line $W$, as in \eref{VpGH}. The low energy gauge
group of this vacuum is the subgroup $S \subset H$ given in
\eref{S}. The particle spectrum is determined as follows:
\begin{itemize}
\item Decompose $\ad E_8$ as in \eref{adE8} in terms of irreducible
	$G$-representations $U_i$ and corresponding
	$H$-representations $R_i$.
\item Decompose each $R_i$ as in \eref{Ri} in terms of irreducible
	$F$-representations $A_j$ and corresponding blocks of
	irreducible $S$-representations $B_{ij}$.
\item Most of the work then goes into computing the cohomology groups
	$H^q(\tilde{X}, U_i(\tilde{V}))$ of the associated vector
	bundles on $\tilde{X}$, and the action of $F$ on these
	cohomologies. The multiplicity $m_{ij}$ of the irreducible
	$F$-representation $A_j^*$ in $H^q(\tilde{X}, U_i(\tilde{V}))$ 
	is the multiplicity of all particles
	from block $B_{ij}$ in the particle spectrum of $(X,\cV')$.
\end{itemize}

We illustrate the general procedure in two cases. First consider
$G = SU(3)$, $H = E_6$. As we saw in \eref{248-3}, the $U_i$ are 1, 8,
3 and $\ol{3}$, and the corresponding $R_i$ are 78, 1, 27 and
$\ol{27}$. Now $H=E_6$ has a maximal subgroup
\beq
H_0 = SU(3)_C \times SU(3)_L \times SU(3)_R \ ,
\eeq
where we can think of $C$, $L$, $R$ as standing for color, left,
right. 
We can, for example, take $F = F(n,\hat{n}) = \IZ_n \times
\IZ_{\hat{n}}$ whose two generators are mapped to $H_0$ as
\beq
\II_C \times 
\mat{\alpha & & \cr &\alpha& \cr &&\alpha^{-2}}_L \times \II_R \ ,
\qquad \qquad
\II_C \times \II_L \times 
\mat{\hat{\alpha} & & \cr &\hat{\alpha}& \cr &&\hat{\alpha}^{-2}}_R 
\ ,
\eeq
where $\alpha$ and $\hat{\alpha}$ are roots of unity of orders $n$ and
$\hat{n}$ respectively. Another possibility is to work with $F_0$, the
diagonal subgroup $\IZ_n$ in $F(n,n)$, with generator
\beq
\II_C \times 
\mat{\alpha & & \cr &\alpha& \cr &&\alpha^{-2}}_L \times 
\mat{\alpha & & \cr &\alpha& \cr &&\alpha^{-2}}_R \ .
\eeq
Either $F$ (with $n, \hat{n} > 1$) or $F_0$ (with $n > 1$) break $E_6$ to
\beq
S = SU(3)_C \times \left(\frac{SU(2) \times U(1)}{\IZ_2}\right)_L
\times \left(\frac{SU(2) \times U(1)}{\IZ_2}\right)_R \ .
\eeq
In this case, it is easier to first decompose each $R_i$ under $H_0$,
and then to further decompose each $H_0$ component under $F$ and
$S$. Under $H_0$ we have:
\bea
78 &=& (8,1,1) \oplus (1,8,1) \oplus (1,1,8) \oplus (3,3,3) \oplus
	(\ol{3},\ol{3},\ol{3}) \nn \\
1 &=& (1,1,1) \nn \\
27 &=& (3,\ol{3},1) \oplus (1,3,\ol{3}) \oplus (\ol{3},1,3) \nn \\
\ol{27} &=& (\ol{3},3,1) \oplus (1,\ol{3},3) \oplus (3,1,\ol{3}) \ ,
\eea
where $(a,b,c)$ is shorthand for the $H_0$-representation $a_C \otimes
b_L \otimes c_R$. When we further decompose under $S$, the color
representations are unchanged, while the 3 of $L$ or $R$ breaks as
$2_1 \oplus 1_{-2}$, and the adjoint 8 breaks as $3_0 \oplus 1_0 \oplus
2_3 \oplus 2_{-3}$. (Here $b_w$ denotes the $b$-dimensional
representation of $SU(2)$, on which $U(1)$ acts with weight $w$. This
representation of $SU(2) \times U(1)$ factors through $(SU(2) \times
U(1))/\IZ_2$ if and only if the integers $b$ and $w$ have opposite
parity.) So the $(8,1,1)$ of $H_0$ becomes $(8,1,1)_{0,0}$ of $S$,
while the $(1,8,1)$ becomes $(1,3,1)_{0,0} \oplus (1,1,1)_{0,0} \oplus
(1,2,1)_{3,0} \oplus (1,2,1)_{-3,0}$. The two subscripts give the
weights of the two $U(1)$'s in $S$. The same subscripts taken modulo
$n$ and $\hat{n}$ give the weights of $F(n,\hat{n})$, so they
determine the representation $A_j$. 
We tabulate the results in \tref{tab:eg1}. 
In that table, the only reducible block is $B_{00}$. However, if
we replace $F$ by its subgroup $F_0$, many of the $A_j$ coalesce,
resulting in many reducible $B_{ij}$'s.

\begin{table}
\[
\ba{||c|c|c|c|c||}\hline\hline
U_i & H^q(\tilde{X}, U_i(\tilde{V})) & R_i & A_j & B_{ij} \\ \hline
\hline
1 & H^0(\tilde{X}, \cO_{\tx}) & 78 & 0,0 & 
	(8,1,1)\oplus(1,3,1)\oplus(1,1,3)\oplus2 \times (1,1,1) \\
	\hline
& & & 3,0 & (1,2,1) \\ \hline
& & & -3,0 & (1,2,1) \\ \hline
& & & 0,3 & (1,1,2) \\ \hline
& & & 0,-3 & (1,1,2) \\ \hline
& & & 1,1 & (3,2,2) \\ \hline
& & & -2,-2 & (3,1,1) \\ \hline
& & & -1,-1 & (\ol{3},2,2) \\ \hline
& & & 2,2 & (\ol{3},1,1) \\ \hline
& & & 1,-2 & (3,2,1) \\ \hline
& & & -2,1 & (3,1,2) \\ \hline
& & & -1,2 & (\ol{3},2,1) \\ \hline
& & & 2,-1 & (\ol{3},1,2) \\ \hline
8 & H^1(\tilde{X}, \ad \tilde{V}) & 1 & 0,0 & (1,1,1) \\ \hline
3 & H^1(\tilde{X},\tilde{V}) & 27 & -1,0 & (3,2,1) \\ \hline
& & & 2,0 & (3,1,1) \\ \hline
& & & 1,-1 & (1,2,2) \\ \hline
& & & -2,-1 & (1,1,2) \\ \hline
& & & 1,2 & (1,2,1) \\ \hline
& & & -2,2 & (1,1,1) \\ \hline
& & & 0,1 & (\ol{3},1,2) \\ \hline
& & & 0,-2 & (\ol 3,1,1) \\ \hline
\ol 3 & H^1(\tilde{X},\tilde{V^*}) & 
	\ol{27} & -1,0 & (\ol 3,2,1) \\ \hline
& & & 2,0 & (\ol 3,1,1) \\ \hline
& & & 1,-1 & (1,2,2) \\ \hline
& & & -2,-1 & (1,1,2) \\ \hline
& & & 1,2 & (1,2,1) \\ \hline
& & & -2,2 & (1,1,1) \\ \hline
& & & 0,1 & (3,1,2) \\ \hline
& & & 0,-2 & (3,1,1) \\ \hline \hline
\ea
\]
\caption{The decomposition of $H^q(X, \ad \cV')$ where $G=SU(3)$
and $F=\IZ_n \times \IZ_{\hat{n}}$.}\label{tab:eg1}
\end{table}

For our second example we consider $G=SU(5)$, so $H=SU(5)$ and the
decomposition of $\ad E_8$ is given in \eref{E8-su5}. The finite group
$F$ is $\IZ_2$, where the generator is embedded in $H=SU(5)$
diagonally with eigenvalues $(1,1,1,-1,-1)$. This breaks $H$ to the
standard model group $S = \sm$. In
\tref{tab:eg2}, we use $(a,b)_w$ to denote the product of an
$a$-dimensional representation of $SU(3)$ with a $b$-dimensional
representation of $SU(2)$, where $U(1)$ acts with weight $w=3Y$. The
corresponding representation $A_j$ of $F$ depends only on the parity
of $w$.
\begin{table}[h]
\[
\ba{||c|c|c|c|c||}\hline\hline
U_i & H^q(\tilde{X}, U_i(\tilde{V})) & R_i & A_j & B_{ij} \\ \hline
\hline
1 & H^0(\tilde{X}, \cO_{\tx}) & 24 & 0 & 
	(8,1)_0\oplus(1,3)_0\oplus(1,1)_0 \\ \hline
	& & & 1 & (3,2)_{-5}\oplus(\ol 3,2)_{5} \\ \hline
24 & H^1(\tilde{X}, \ad \tilde{V}) & 1 & 0 & (1,1)_0 \\ \hline
10 & H^1(\tilde{X}, \avt) & 5 & 0 & (3,1)_{-2} \\ \hline
	& & & 1 & (1,2)_{3} \\ \hline
\ol{10} & H^1(\tilde{X}, \avt^*) & \ol 5 & 0 & (\ol 3,1)_{2} \\ \hline
	& & & 1 & (1,2)_{-3} \\ \hline
5 & H^1(\tilde{X}, \tv) & \ol{10} & 0 & (3,1)_{4} \oplus (1,1)_{-6} 
	\\ \hline
	& & & 1 & (\ol 3,2)_{-1} \\ \hline
\ol 5 & H^1(\tilde{X}, \tv^*) & 
	10 & 0 & (\ol 3,1)_{-4} \oplus (1,1)_{6} \\ \hline
	& & & 1 & (3,2)_{1} \\ \hline \hline
\ea
\]
\caption{The decomposition of $H^q(X, \ad \cV')$ where $G=SU(5)$
and $F=\IZ_2$.
The $A_j$ correspond to characters of the $\IZ_2$ action on $R_i$.
The $a,b$ in $(a,b)_w$ are the representations of $SU(3)_C$ and
$SU(2)_L$ respectively, whereas $w = 3Y$.}\label{tab:eg2}
\end{table}

%%%%%%%%%%%%%%%%
%
	\section{Standard Model Bundles}
%
%%%%%%%%%%%%%%%%%
In this section we recall the standard model bundles constructed in
\cite{Donagi:2000fw,Donagi:2000zs,Donagi:2000zf}. 
We need a quadruple $(\tx, A, \tau, \tv)$ satisfying:
\bea\label{cond}
\bullet&({\bf \IZ_2})& \mbox{$\tx$ is a smooth
	Calabi-Yau 3-fold and
$\tau : \tx \to \tx$ is a freely acting involution.}\nn\\ 
	&&\mbox{$A$ is a fixed
	ample line bundle (K\"{a}hler structure) on $\tx$.}\nn\\
\bullet&{\bf (S)}& \mbox{$\tv$ is an $A$-stable vector bundle of rank
five on $\tx$ with structure group $G=SU(5)$.}\nn\\
\bullet&{\bf (I)}& \mbox{$\tv$ is $\tau$-invariant.}\nn\\
\bullet&{\bf (C1)}& \mbox{$c_{1}(\tv) = 0$.}\nn\\
\bullet&{\bf (C2)}& \mbox{$c_{2}(\tx) - c_{2}(\tv)$ is effective.}\nn\\
\bullet&{\bf (C3)}& \mbox{$c_{3}(\tv) = 12$.}
\eea
The involution $\tau$ generates a subgroup
$\IZ_2 = F \subset H = Z_{E_8}(SU(5)) = SU(5)$.
%and the associated Wilson line further reduces $H$ to the standard
%model gauge group $S = \sm$.
The quotient $X := \tx / F$ is another
Calabi-Yau threefold, and invariance of $\tv$ allows us to identify it
with the pullback of a stable $SU(5)$ bundle $V$ on $X$, as in
\sref{s:1}. 
%Combining this with the $\IZ_2$ Wilson line $W$ on $X$
This produces a heterotic M-theory vacuum $(X,\cV')$ with
particle spectrum as given in \tref{tab:eg2} of \sref{s:1}.

%%%%%%%%-----------------------
\subsection{Rational Elliptic Surfaces and Their Products}
The simply connected threefold $\tx$ is a complete intersection in
$\IP^1 \times \IP^2 \times (\IP^2)'$ of two hypersurfaces of
multidegrees $(1,3,0)$ and $(1,0,3)$ respectively. This is a
Calabi-Yau, by adjunction, and it has two
elliptic fibrations. These threefolds were first studied by Schoen
\cite{schoen}. Choose projective coordinates: $[t_0 : t_1]$ on $\IP^1$; 
$z = [z_0 : z_1 : z_2]$ on $\IP^2$; and $z' = [z_0' : z_1' : z_2']$ on
$(\IP^2)'$. The two hypersurfaces can be written:
\bea
t_0 f_0(z) - t_1 f_1(z) &=& 0 \label{h1} \\
t_0 f_0'(z') - t_1 f_1'(z') &=& 0 \label{h2} \ ,
\eea
where $f_0, f_1, f_0', f_1'$ are homogeneous cubic polynomials. Since
equation \eref{h1} does not involve $z'$, it defines a hypersurface $B
\subset \IP^1 \times \IP^2$. Similarly equation \eref{h2} defines a
hypersurface $B' \subset \IP^1 \times (\IP^2)'$. The surfaces $B$,
$B'$ are called rational elliptic surfaces, or
(inaccurately) $dP_9$'s.
The projections of these surfaces to $\IP^1$ are elliptic fibrations:
\beq
\beta : B \to \IP^1, \qquad \beta' : B' \to \IP^1 \ .
\eeq
The original threefold $\tx$ comes with the two projections
\beq\label{pipip}
\pi : \tx \to B', \qquad \pi' : \tx \to B
\eeq
which are again elliptic fibrations. In fact, $\tx$ is the fiber
product
\beq\label{fiberprod}
\tx = B \times_{\IP^1} B' \ ,
\eeq
meaning that a point of $\tx$ is uniquely specified by a pair of
points $b \in B$, $b' \in B'$ with $\beta(b) = \beta'(b') \in
\IP^1$.

The opposite projection $\nu : B \to \IP^2$ is birational,
exhibiting $B$ as the blowup of $\IP^2$ at the 9 points $A_i$, $i = 1,
\ldots, 9$ where $f_0(z) = f_1(z)=0$, and similarly for $B'$. 
(This is the origin of the ``$dP_9$'' name -- but these surfaces are
not del Pezzos.)
It follows that $H^2(B, \IZ) = Pic(B)$ has rank 10. An orthogonal
basis consists of the class $\ell := \nu^* \cO_{\IP^2}(1)$ together
with the 9 exceptional classes $e_1, \ldots, e_9$. The only non-zero
intersection numbers on $B$ are
$\ell^2 = 1, \quad e_i^2 = -1$, $i = 1, \ldots, 9$. 
The class $f := \beta^{-1}(\mbox{point})$ of an elliptic fiber is given
by $f = 3 \ell - \sum\limits_{i=1}^9 e_i$. 
There is an analogous basis
$\ell', e_1', \ldots, e_9'$ on $B'$. 
%Using \eref{pipip} and
%\eref{fiberprod}, it follows that 
The rank of $H^2(\tx, \IZ)$ is
therefore 19, with basis $\pi^*\ell' = (\pi')^*\ell, \pi^*e_1', \ldots,
\pi^*e_9', (\pi')^*e_1, \ldots, (\pi')^*e_9$.

%%%%%%%%%%%%%%%
\subsection{Special Rational Elliptic Surfaces}\label{s:B}
In order to obtain the involution $\tau$ on $\tx$, and also in order
to have invariant bundles $\tv$ on $\tx$ satisfying the required
conditions, the rational elliptic surfaces $B$, $B'$ need to be
specialized to a particular subfamily. This can be specified as
follows.

Let $\Gamma_{1} \subset \IP^{2}$ be a nodal cubic with a node
$A_{8}$. Choose four generic points on $\Gamma_{1}$ and label them
$A_{1}, A_{2}, A_{3}, A_{7}$. Let $\Gamma \subset \IP^{2}$ be the
unique smooth cubic which passes through $A_{1}, A_{2}, A_{3}, A_{7},
A_{8}$ and is tangent to the lines $\langle A_{7}A_{i} \rangle$ for $i
= 1, 2, 3$ and $8$. Consider the pencil of cubics spanned by
$\Gamma_{1}$ and $\Gamma$. All cubics in this pencil pass through 
$A_{1}, A_{2}, A_{3}, A_{7}, A_{8}$ and are tangent to $\Gamma$ at
$A_{8}$. Let $A_{4}, A_{5}, A_{6}$ be the remaining three base points,
and let $B$ denote the blow-up of $\IP^{2}$ at the points $A_{i}$, $i =
1,2, \ldots, 8$ and the point $A_{9}$ which is infinitesimally near
$A_{8}$ and corresponds to the line $\langle A_{7}A_{8} \rangle$.

The pencil becomes the anti-canonical map $\beta : B \to \IP^{1}$ which
is an elliptic fibration with a section. The map $\beta$ has two
reducible fibers $f_{i} = n_{i}\cup o_{i}$, $i = 1,2$ of type
$I_{2}$. We denote by $e_{i}$, $i = 1, \ldots, 7$ and $e_{9}$ the
exceptional divisors corresponding to $A_{i}$, $i = 1, \ldots, 7$ and
$A_{9}$, and by $e_{8}$ the reducible divisor $e_{9} + n_{1}$. The
divisors $e_{i}$ together with the pullback $\ell$ of a class of a line
from $\IP^{2}$ form a standard basis  in $H^{2}(B,{\mathbb Z})$.

The surface $B$ has an involution $\alpha_{B}$ which is uniquely
characterized by the properties: $\b \circ \alpha_{B} = \tau_{\IP^1}
\circ \b$, where $\tau_{\IP^1}$ is the involution
$t_0 \to t_0, t_1 \to -t_1$ on $\IP^{1}$,
and $\alpha_{B}$ fixes
the proper transform of $\Gamma$ pointwise. Note that $\tau_{\IP^1}$
leaves two points in $\IP^1$ fixed, which we call $0$ and
$\infty$. Furthermore, $\alpha_B$ acts as $(-1)_B$ when
restricted to the fiber $f_\infty = \b^{-1}(\infty)$ and, hence,
leaves four points fixed in $f_\infty$.

Choosing  $e_{9} := e$ as the zero section of $\beta$, we can interpret
any other section $\xi$ as an automorphism $t_{\xi} : B \to B$
which acts along the fibers of $\beta$. The automorphism $\tau_{B} =
t_{e_1}\circ \alpha_{B}$ is again an involution of $B$ which
commutes with $\beta$, induces the same involution on $\IP^{1}$ as
$\alpha_{B}$, and has four isolated fixed points sitting on the
same fiber $f_\infty$ of $\beta$.

\begin{figure}
   \parbox{\textwidth}{
   \begin{center}
      \psfig{file=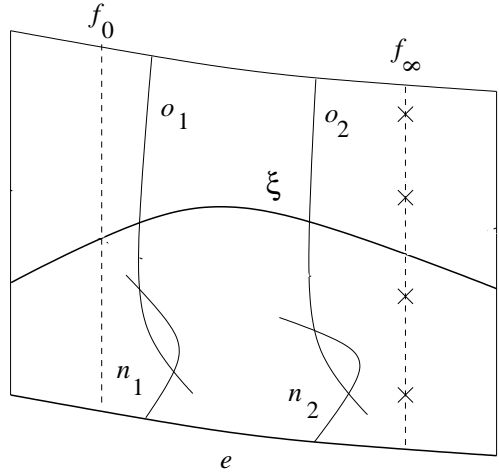,height=2.5in}
   \end{center}
   \caption{A special rational elliptic surface $B$. It has 8 $I_1$
singular fibers. 
In addition, there are 2 $I_2$ fibers
$f_1 = n_1 \cup o_1$ and $f_2 = n_2 \cup o_2$.
Under the involution $\tau_B = t_\xi \circ \alpha_B$, there are 4
fixed points, which we have marked, on the fiber $f_\infty$.}
\label{fig:B}
}
\end{figure}

The special rational elliptic surfaces form a four dimensional
irreducible family. Their geometry was the subject of \cite{Donagi:2000fw}.
The structure of a special rational elliptic surface
$B$ is shown in \fref{fig:B} and the action of $\tau_B$ on 
$H^2(B,\IZ)$ is summarized in \tref{tab:tB}.

\begin{table}
\[
\ba{||l||l||}
\hline\hline
                           & \tau_{B}^{*} \\ 
\hline
        e_1              & e_9 \\
        e_j (j=2,3)    & f - e_j + e_1 + e_9 \\
        e_i (i=4,5,6)  & f - l + e_i + e_1 + e_7 + e_9 \\
        e_7              & l - e_2 - e_3 \\
        e_8              & f - l + e_1 + e_7 + e_8 + e_9 \\
%\hline
%	n_1 = e_8 - e_9  & o_2 \\
%	o_1 = f - n_1 & n_2 \\
%	n_2 = l - e_7 - e_8 - e_9  & o_1 \\
%	o_2 = f - n_2 & n_1 \\
%	f_1 = n_1 \cup o_1 & f_2 \\
%	f_2 = n_2 \cup o_2 & f_1 \\
%\hline
        e_9=e            & e_1 \\
        l                & 2f + 2(e_1+e_9) - (e_2+e_3) + e_7 \\
        f = 3l - \sum\limits_{i=1}^9 e_i & f \\
\hline \hline
\ea
\]
\caption{The action of $\tau_B$ on $H^2(B, \IZ)$.}
\label{tab:tB}
\end{table}

% STRUCTURE OF B

%
%  STRCUTURE OF X
\begin{figure}
   \parbox{\textwidth}{
   \begin{center}
      \psfig{file=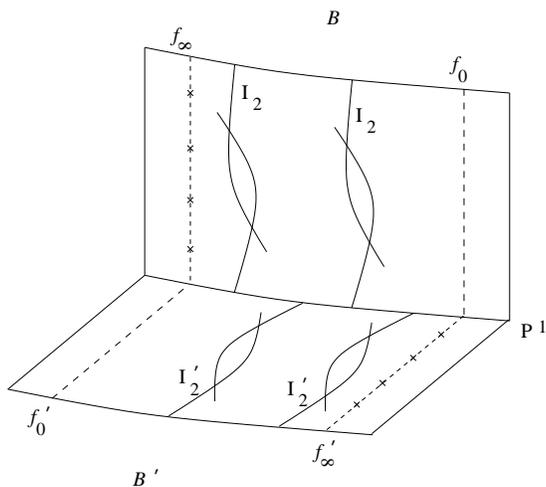,height=2.5in}
   \end{center}
\caption{The Calabi-Yau threefold $\tx$ is constructed as the fiber
product over $\IP^1$
of two non-generic $dP_9$ surfaces $B$ and
$B'$. We have matched the fibers $f_0$ and $f_\infty$ of $B$ with
the fibers $f'_\infty$ and $f'_0$ of $B'$ respectively. The image points
in $\IP^1$ of these fibers, namely $0$ and $\infty$ for $B$ and
$0'$ and $\infty'$ for $B'$, are identified as
$0 = \infty'$ and $\infty = 0'$.}
\label{fig:X}
}
\end{figure}
%

%%%%
\subsection{Building $\tx,\tau$ and $A$}\label{s:tau}
Choose two special rational elliptic surfaces $\beta : B \to \IP^{1}$
and $\beta' : B' \to \IP^{1}$ in $\tx$
so that the discriminant loci of $\beta$
and $\beta'$ in $\IP^{1}$ are disjoint, $\alpha_{B}$ and $\alpha_{B'}$
induce the same involution on $\IP^{1}$, and the fixed loci of $\tau_{B}$
and $\tau_{B'}$ sit over different points 0 and $\infty$
in $\IP^{1}$. The fiber
product $\tx = B\times_{\IP^{1}} B'$ is a smooth Calabi-Yau  threefold
which is elliptic and has a freely acting involution
$
\tau := \tau_{B}\times_{\IP^1}\tau_{B'}
$
and another (non-free) involution $\alpha_{X} := \alpha_{B}\times_{\IP^1}
\alpha_{B'}$.  For concreteness we fix the elliptic
fibration of $\tx$ to 
be the projection
$
\pi : \tx \to B'
$
to $B'$. 
The structure of $\tx$ is shown in \fref{fig:X}.

The stability of the bundle $\tv$ which we describe below is with
respect to a particular choice of K\"ahler class $A$. If $A_0$ is any
K\"ahler class on $\tx$, $h'$ a K\"ahler class on $B'$, and $n \gg 0$,
the class of $A = A_0 + n \pi^* h'$ will be K\"ahler on $\tx$. The
specific value that was found in \cite{Donagi:2000zs} to satisfy all
the requirements was given by
$h' = 193 f' + 144 e_1' + 168(e_9' + e_4' - e_5')$.

%%%%%
\subsection{The Construction of $V$}\label{s:V}
The construction of the $SU(5)$ bundle $V$ on $X := \tx/F$ is
equivalent to the construction of an $SU(5)$ bundle $\tv$ on $\tx$
together with an action of the involution $\tau$ on $\tv$. The
construction of $\tv$ in \cite{Donagi:2000zs} employs a combination of two
techniques: extensions and the spectral construction.

The rank 5 bundle $\tv$ is constructed as an extension
\beq\label{v23}
0 \to V_2 \to \tv \to V_3 \to 0
\eeq
involving two simpler bundles $V_2$, $V_3$, of ranks 2 and 3
respectively. Given the $V_i$, we can immediately construct their
direct sum $\tv_0 = V_2 \oplus V_3$, which is the trivial
extension. In terms of an open cover $\{U_{\alpha}\}$ and $i \times i$
transition matrices $\{g_{i\alpha\beta}\}$ for each $V_i$, the
transition matrices for $\tv_0$ are
\beq
g_{0 \alpha \beta} = 
\mat{g_{2 \alpha \beta} & 0 \cr 0 & g_{3 \alpha \beta}} \ .
\eeq
A general extension $\tv$ is a rank 5 bundle containing $V_2$ as a
subbundle with quotient $V_3$, but $V_3$ cannot be realized as a
subbundle of $\tv$ unless $\tv$ is the trivial extension $\tv_0$. The
transition matrices for such an extension must be of the form:
\beq
g_{\alpha \beta} = 
\mat{g_{2 \alpha \beta} & h_{\alpha \beta} \cr 
	0 & g_{3 \alpha \beta}} \ .
\eeq
In order for these $g_{\alpha \beta}$ to define a vector bundle, the
upper right corner $h_{\alpha \beta}$ must satisfy a cocycle
condition. Working this out shows that the set of isomorphism classes
of extensions is described by the sheaf cohomology group:
\beq
\ext^1_{\tx}(V_3, V_2):= H^1(\tx, V_3^* \otimes V_2) \ .
\eeq
The direct sum $\tv_0 = V_2 \oplus V_3$ corresponds to the 0 element
of this extension group. Our standard model bundle $\tv$ turns out to
correspond to a non-trivial extension $[\tv] \in \ext_{\tx}^1(V_3, V_2)$. In
order for $\tv$ to be $\tau$-invariant, we require first that $V_2$
and $V_3$ be $\tau$-invariant, so we can choose an action of $\tau$ on
$V_2$ and $V_3$. This induces an action of $\tau$ on 
$\ext_{\tx}^1(V_3, V_2)$.
In order for $\tv$ to be $\tau$-invariant, we
require that the extension class $[\tv]$ be $\tau$-invariant.

%We will always take $[\tv]$ to be in $\ext^1_{\tx}(V_3,V_2)_{(+)}$, so
%that the $\tau$-action on $\tv$ is compatible with the actions on
%$V_2$ and $V_3$.
%The opposite choice gives no extra freedom:
%it simply amounts to a $\tau$-action on $\tv$ which is anti-compatible
%with the action on one of the $V_i$, and it can be converted to the
%previous case by flipping our choice of $\tau$-action on the offending
%$V_i$.

%%%%%%%%%%%%%%%%%%%%%
\subsection{The Construction of the $V_i$}
The construction of the bundles $V_i$, $i = 2,3$, involves the
{\bf spectral construction} or {\bf Fourier-Mukai transform}
\cite{FMW1,D,FMW2}. The Fourier-Mukai transform is a self-equivalence
of the derived category $D^b(\tx)$ of coherent sheaves on $\tx$
\bea
FM : && D^b(\tx) \to D^b(\tx) \nn \\
&& \cF \to Rp_{1*}(p_2^* \cF \stackrel{L}{\otimes} \cP) \ .
\eea
Here, $p_1$, $p_2$ are the projections of the fiber product $\tx
\times_{B'} \tx$ to the two $\tx$ factors, $Rp_{1*}$ is the right
derived functor of $p_{1*}$, $\cP$ is the Poincar\'e
sheaf on $\tx \times_{B'} \tx$, and $\stackrel{L}{\otimes}$ is the left
derived functor of $\otimes$.
If $V_i$ is a rank $i$ vector bundle
on $\tx$ which is semistable and of degree 0 on each elliptic fiber
$f$ of $\pi:\tx \to B'$, then $FM^{-1}(V_i)$ is a rank 1 sheaf
$N_{\Sigma_i}$ supported on a divisor $\Sigma_i \subset \tx$ which is
finite of
degree $i$ over the base $B'$. In other words, $\Sigma_i$ intersects
each elliptic fiber $f$ in $i$ points. In case $\Sigma_i$ is smooth, 
$N_{\Sigma_i}$ is actually a line bundle on $\Sigma_i$. The spectral
construction starts with $(\Sigma_i, N_{\Sigma_i})$ and recovers the
bundle $V_i$ as the Fourier-Mukai transform. 
When $\Sigma_i$ is irreducible, the
resulting bundle $V_i$ is automatically stable.

\begin{figure}
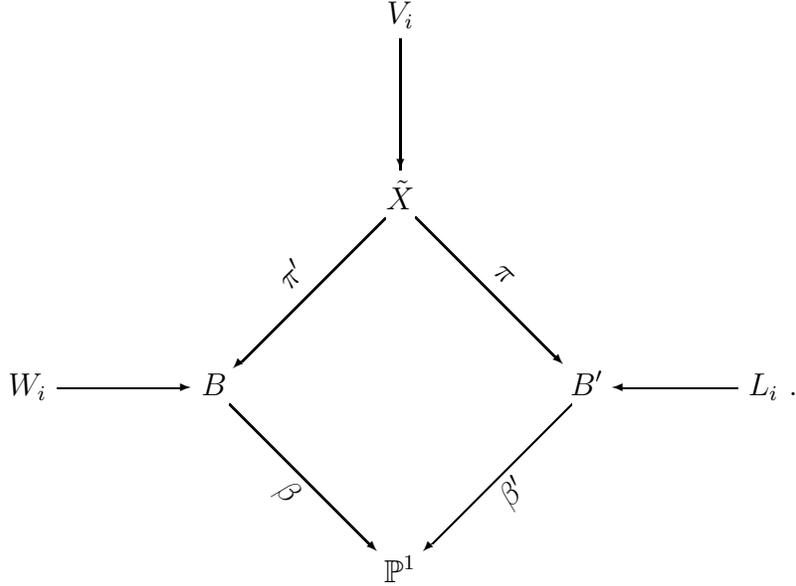

\begin{diagram}
	&&&		&	V_i	&&	\\
	&&&		&	\dTo	&&	\\
	&&&		&	\tx	&&	\\
	&&&\ldTo^{\pi'}	&		& \rdTo{\pi}&	\\
W_i &\rTo &B&		&		&& B'	&\lTo &L_i \ .	&\\
	&&&\rdTo_{\beta}	&		& \ldTo_{\beta'}	&\\
	&&&		&	\IP^1	&&	
\end{diagram}
\caption{The structure of the vector bundles $V_i$, $i=2,3$.}
\label{f:comm}
\end{figure}

In our case we do not need the full spectral construction on the
threefold $\tx$. The map $\beta : B \to \IP^1$ is an elliptic
fibration, so there is a Fourier-Mukai transform $FM_B$ on
$D^b(B)$. We will describe below certain curves $C_i \subset B$ and
line bundles $N_i \in Pic(C_i)$ for $i =2,3$. These determine two
bundles $W_i := FM_B(C_i, N_i)$ with $\rk(W_i) = i$.
Our desired bundles $V_i$
are then recovered as
\beq\label{vi}
V_i = \pi'^* W_i \otimes \pi^* L_i
\eeq
for appropriate line bundles $L_i \in Pic(B')$. The spectral data on
$B$ and on $\tx$ are related by
\beq
\Sigma_i = (\pi')^{-1} C_i = C_i \times_{\IP^1} B', \qquad
N_{\Sigma_i} = (\pi')^* N_i \otimes \pi^* L_i \ .
\eeq
This is summarized in \fref{f:comm}.

The specific values we take for the $C_i$, $N_i$ and $L_i$ are as
follows. Choose curves $\overline{C}_{2}, C_{3} \subset B$, so that
\[
\ba{lcl}
\bullet&& \overline{C}_{2} \in |{\mathcal O}_{B}(2e_{9} + 2f)|, \quad
	C_{3} \in |{\mathcal O}_{B}(3e_{9} + 6 f)|, \nn \\ 
\bullet&& \mbox{$\overline{C}_{2}$ and $C_{3}$ are
	$\alpha_{B}$-invariant,} \nn\\
\bullet&& \mbox{$\overline{C}_{2}$ and $C_{3}$ are smooth and irreducible.}
\ea
\]
Set $C_{2} =
\overline{C}_{2} + f_{\infty}$ where $f_{\infty}$ is the smooth fiber
of $\beta$ containing the four fixed points of $\tau_{B}$. 
We choose the line bundles $N_2 \in Pic^{3,1}(C_2)$, $N_3 \in Pic^{16}(C_3)$
to transform correctly under the involution $\alpha_B|_{C_i}$:
\beq\label{Ni}
N_i \simeq 
(\alpha_B |_{C_i})^* N_i \otimes \cO_{C_i}(e_1 -e_9+f), \qquad
i = 2,3 \ .
\eeq
Here $Pic^{3,1}(C_2)$ denotes line bundles of degree 3 on
$\ol{C}_2$ and degree 1 on $f_\infty$ \cite{Donagi:2000zs}.
(It is shown in \cite{Donagi:2000zs} that such $N_i$ do exist.)
%The Chern character of $W_i$ is given by
%\beq\label{chernWi}
%\ch(W_i) = r_i + (10 + \left( \ba{c} r_i+1\\ 2\ea\right) - k_i r_i
%-r_i) f - k_i \pt,
%\eeq
%where
%\beq\label{chernWirk}
%(r_2, k_2) = (2, 3), \quad
%(r_3, k_3) = (3, 6).
%\eeq
%and $d_{2,3}$ are some arbitrary integer parameters such that $d_2$
%is even while $d_3$ is 1 modulo 3.
A useful quantity associated with the bundle $W_2$
is the degree $-1$ line bundle $G \in Pic^{-1}(f_\infty)$ on
the elliptic curve $f_\infty$, defined as
\beq\label{defG}
G = N_2 |_{f_\infty}(-D),
\eeq
where $D$ is the divisor $D = \ol{C}_2 \cap f_\infty$.
%In particular,
%\beq\label{degG}
%\deg G = -1.
%\eeq
This fits into an exact sequence
\beq\label{seqW2}
0 \to W_2 \to \ol{W}_2 \to i_{f_\infty *}(G^*) \to 0 \ ,
\eeq
where $\ol{W}_2$ is the rank 2 vector bundle associated
with the spectral cover $\ol{C}_2$ and spectral line bundle
$\ol{N}_2 = N_2 \otimes \cO_{\ol{C}_2}$. 
The Chern characters can be read from Lemma 5.1 of \cite{Donagi:2000zs}: 
\beq\label{chernWi}
\ba{cc}
\ch(W_2) = 2 - f - 3 \pt, &
\ch(\ol{W}_2) = 2 - 2 \pt, \nn \\
\ch(W_3) = 3 + f - 6 \pt, &
\ch(G^*) = f + \pt.
\ea
\eeq
%It follows from this and
%\eref{chernWi} that
%\beq\label{chernW2}
%\ch(\ol{W}_2) = 2 + 6 f - 2 \pt.
%\eeq

%The Lemma states that the first chern class of a bundle $W$ on $B$, 
%coming from the Fourier-Mukai transform
%of the spectral data $(C \in |\cO_B(r e_9 + m f)|, N \in Pic^d(C))$, is:
%\beq
%c_1(W) = d + \left(\ba{c} r+1\\ 2 \ea \right) - r m - r.
%\eeq
%Now, $\ol{W}_2$ has $r_2=m_2=2$, so $d_2$ is chosen to be 3, to give
%the Chern class in \eref{chernWi}.
%Similarly, $W_3$ has $r_3=3$, $m_3=6$ and so $d_3$ is chosen to be
%16. This is why in the previous paragraph, we claimed that
%$N_2 \in Pic^{3,1}(C_2)$, $N_3 \in Pic^{16}(C_3)$.

Finally, the line bundles $L_i$ on $B'$ are given by
%\bea\label{L23}
%L_2 &=& 3(e_1' + e_4' - e_5' + e_9') + f' + 3(n_1' + o_2') \nn \\
%L_3 &=& -2(e_1' + e_4' - e_5' + e_9') + 5f' -2(n_1' + o_2') \ .
%\eea
\bea\label{L23}
L_2 &=& \obp(3 r') \nn \\
L_3 &=& \obp(-2 r')
\eea
where
\beq\label{defr}
r' = e_1' + e_4' - e_5' + e_9' + f'
= 3\ell' - 2e_4' - (e_2'+e_3'+e_6'+e_7'+e_8').
\eeq
Formula \eref{L23} holds with the specific choices $N_2 \in Pic^{3,1}(C_2)$,
$N_3 \in Pic^{16}(C_3)$ which we made above, and only with those
choices. This is why we specify the general solution in \cite{Donagi:2000zs}
to these values.

This completes the specification of the bundles
$V_i$ for $i=2,3$.
%The constraint that $\bar{C}_2$, $C_3$ be
%$\alpha_B$-invariant, together with
%\eref{Ni} and \eref{L23}, guarantee that these
%bundles are $\tau$-invariant.
It was seen in \cite{Donagi:2000zs} that $\tau$-invariant extensions
$[\tv] \in \ext_{\tx}^1(V_3, V_2)_{(+)}$ exist,
and that the bundle $\tv$ corresponding
to a general such $[\tv]$ has structure group $G = SU(5)$, is stable, is
$\tau$-invariant, and satisfies the requirements $(\IZ_2, S, I, C1, C2,
C3)$ in \eref{cond}.

%%%%%%%
\subsection{Comments}
The reason we did not build $\tv$ directly by a spectral construction
applied to the surface $\Sigma = \Sigma_2 \cup \Sigma_3$ in $\tx$  (or
to the curve $C = C_2 \cup C_3$ in $B$) is that on singular spectral
covers (such as $\Sigma$, $C$), the rank 1 sheaf ($\ol{N}$ or $N$) can
fail to be a line bundle, leading to technical complications. A
closely related problem is that it is harder to check the stability of
$\tv$ when the spectral cover is reducible.

Another subtlety is that our $C_2$ is not finite over $\IP^1$. It
intersects the generic elliptic fiber in 2 points, but it contains the
entire fiber $f_\infty$. 
%Therefore $C_2$, and similarly $\Sigma_2$, cannot
%be used to obtain a rank 2 bundle $\ol{W}_2$. The desired $W_2$, whose
%inverse Fourier-Mukai transform is supported on $C_2$, is related to
%this $\ol{W}_2$ by the exact sequence \eref{seqW2}.
We chose $N_2$ carefully so that our $W_2$ is still the Fourier-Mukai
transform of $(C_2,N_2)$.
But in practice it is often easier to work with $\ol{C}_2$, $\ol{N}_2$
and $\ol{W}_2$, and to relate $W_2$ and $\ol{W}_2$ via \eref{seqW2}.

The construction in \cite{Donagi:2000zs} involves additional degrees of
freedom in the form of Hecke transforms applied to the $\tv$. Later
checks, motivated by questions of Mike Douglas,
suggest that most or
all of these extra degrees of freedom
%namely the $a_{\kappa}$ and
%$d_i$ coefficients in \cite{Donagi:2000fw}, 
may be illusory. 
At any rate, we do not use them in the present work.

%%%%%%%%%%%%%%%%%%%%%%%%%%%%%%%%%%%%%%%%%%%%
%%%%%%
	\section{Cohomologies of $U_i(\tv)$}\label{s:coh}
%%%%%
%%%%%%%%%%%%%%%%%%%%%%%%%%%%%%%%%%%%%%%%%%%%%
In order to compute the relevant cohomologies on a rational elliptic
surface such as $B'$, we need some basic facts about the line bundle
$\obp(r')$ of \eref{defr}.
We claim that the direct image is:
\beq\label{br}
\b'_* \obp(r') \simeq \op1 \oplus \op1,
\eeq
or equivalently that
\beq\label{b(r-f)}
\b'_* \obp(r'-f') \simeq \op1(-1) \oplus \op1(-1).
\eeq
Indeed, the left hand side of \eref{b(r-f)} is a rank 2 bundle on
$\IP^1$, since $(r'-f') \cdot f' = 2$, so it must be of the form
$\op1(a) \oplus \op1(b)$ for some integers $a$, $b$.
Now $r'-f' = e_1' + e_9' + e_4' - e_5'$ cannot be effective 
(any effective representative has negative intersection with $e_1'$,
$e_4'$, $e_9'$, so must contain all of them), and therefore our
integers $a$, $b$ must be negative.
To conclude that $a=b=-1$ as claimed in \eref{b(r-f)}, it suffices to
note that $a+b$ is the degree of $\b'_* \obp(r'-f')$, which by 
\GRR~(GRR) equals $-2$.

Instead of GRR, we can obtain the same result using a bit of geometry.
We saw in \eref{defr} that $r' = 3\ell' -(e_2'+e_3'+e_6'+e_7'+e_8') -
2e_4'$, so we can identify sections of $\obp(r')$
with cubic polynomials on $\IP^2$ vanishing at $A_i$ for
$i=2,3,6,7,8$, and vanishing to second order at $A_4$.
The space $H^0(\cO_{\IP^2}(3 \ell))$ of cubics is
10 dimensional, the vanishing at each of the five $A_i$ imposes one
linear condition, and vanishing to second order at $A_4$ imposes 3
more conditions, for a total of 8 conditions.
Therefore $2= 10-8 \le h^0(\obp(r')) = 
h^0(\IP^1,\b'_* \obp(r')) =
h^0(\op1(a+1)) + h^0(\op1(b+1))$.
Recalling that $a$, $b$ are negative, this is possible only for
$a=b=-1$; so we have found another argument for \eref{br},
\eref{b(r-f)}.

It follows from \eref{br} that $H^0(\obp(r'))$ is 2 dimensional.
We let $x_0$ and $x_1$ be a basis.
We claim that the quotient $x_1 / x_0$ is everywhere defined, so it
gives a map 
\beq
\chi : B' \to \IP^1_x,
\eeq
and the $x_i$ can be interpreted as homogeneous coordinates on the
target $\IP^1_x$.
Checking that $\chi$ is everywhere defined is equivalent to verifying
that $x_0$ and $x_1$ cannot vanish at the same point.
Since $r'^2 = 0$, two divisors in the linear system $|r'|$ cannot
intersect each other unless they have a common component.
So to conclude, it suffices to check that some (and hence almost
all) of these divisors are irreducible.
This follows immediately from the geometric model: in fact, the fibers
of $\chi$, identified as the pencil of cubics vanishing at the five
$e_i'$ and doubly at $e_4'$, include precisely 8 reducible curves,
namely:
\beq\label{Kij}
\ba{lll}
K_i^1 \cup K_i^2, &
K_i^1 = \ell' - e_5' - e_i', &
K_i^2 = 2\ell'-(e_2'+e_3'+e_6'+e_7'+e_8')-e_5'+e_i',
	\quad i=2,3,6,7,8 \\
K_j^0 \cup K_j^3, &
K_j^0 = e_j', &
K_j^3 = 3\ell'-(e_2'+e_3'+e_6'+e_7'+e_8')-2e_5'-e_j',
	\quad j=1,4,9.
\ea\eeq
The first five curves occur as reducible cubics in $\IP^2$, consisting
of the line joining $A_5$ to $A_i$ and the conic through $A_5$ and the
remaining 4 points.
The last three consist of cubics which happen to pass through one of
the $A_j$, so their inverse image in $B'$ contains the corresponding
$e_j'$.
All other cubics in our system are singular (at $A_5$) but
irreducible.
We conclude that $\chi$ is indeed everywhere defined, its generic
fiber is a $\IP^1$, and precisely the 8 fibers listed in \eref{Kij}
split into a pair of $\IP^1$'s meeting at one point.

Clearly, the target space $\IP^1_x$ of the map $\chi$ defined by the
line bundle $\obp(r')$ has nothing to
do with the target space $\IP^1_t$ of the map $\b'$ defined by the
line bundle $\obp(f')$.
In fact, we can put these two maps together, to get a map
\beq
\Delta = (\b', \chi) : B' \to \q := \IP^1_t \times \IP^1_x
\eeq
given by the two pairs of homogeneous coordinates $(t_0, t_1)$,
$(x_0, x_1)$.

The product surface $\q$ could be identified with a smooth quardric in
$\IP^3$ via the embedding $(t_0x_0, t_1x_0, t_0x_1, t_1x_1)$, but we
will not use this.
The product map $\Delta$ is onto $\q$, and is of degree $f' \cdot r' =
2$; in other words, we have realized the rational elliptic surface
$B'$ as a double cover of the quadric surface $\q$.
The fibers of $\b'$ are of course the elliptic curves $f'$ which now
appear as double covers of $\IP^1_x$ branched at 4 points.
The general fiber of $\chi$, on the other hand, is isomorphic to a
$\IP^1$, as is seen by adjunction.
It appears as a double cover of $\IP^1_t$ branched at 2 points.
The branch locus $Br_\Delta$ of $\Delta$ is therefore a divisor of
bidegree $(4,2)$ in \q.

Line bundles on \q are of the form $\cO_{\q}(k,l) :=
pr_t^*\cO_{\IP^1_t}(l) \otimes pr_x^*\cO_{\IP^1_x}(k)$, with integers
$k$, $l$, where $pr_t$, $pr_x$ are the projections to $\IP^1_t$,
$\IP^1_x$ respectively: $\b' = pr_t \circ \Delta$, $\chi = pr_x
\circ \Delta$.
Let us introduce the abbreviation
\beq
\obp(k,l) := \Delta^* \cO_{\q}(k,l) = \obp(k r' + l f')
\eeq
for the corresponding line bundles on $B'$.
So for example $\obp(0,1)$ is the anticanonical bundle $K_{B'}^{-1}
\simeq \obp(f')$, $\obp(1,0)$ is $\obp(r')$, $\obp(3,0)$ is our $L_2$,
and $\obp(-2,0)$ is $L_3$.

On $B'$ there is a unique involution $\iota$ which exchanges the two
sheets of $B'$ over \q.
Its fixed locus is the ramification divisor $Ram_\Delta \subset B'$.
The image $\Delta(Ram_\Delta)$ is of course $Br_\Delta$.
Since
\beq
\Delta^*\cO_{\q}(Br_\Delta) = \obp(2Ram_\Delta)
\eeq
and the Picard group of $B'$ has no torsion, we find that:
\beq\label{ob-ram}
\obp(Ram_\Delta) \simeq \Delta^*\cO_{\q}\left(\frac12 Br_\Delta\right) =
\Delta^*\cO_{\q}(2,1) = \obp(2,1).
\eeq
For any double cover such as $\Delta$, sections of $\obp$ can be
decomposed into $\iota$-invariants and anti-invariants.
This can be written formally as a decomposition of the direct image:
\beq
\Delta_* \obp \simeq 1\cdot \cO_{\q} \oplus y \cdot 
	\cO_{\q}\left(-\frac12 Br_\Delta\right),
\eeq
where $y \in H^0(\obp(Ram_\Delta))$ is the
$\iota$-anti-invariant
section characterized up to scalars by its vanishing precisely on
$Ram_\Delta$. (This is another special case of GRR).
In our case, \eref{ob-ram} shows that
\beq\label{def-y}
y \in H^0(\obp(2,1)), \quad \iota y = -y
\eeq
and
\beq
\Delta_* \obp = \cO_{\q} \oplus y \cO_{\q}(-2,-1).
\eeq
This can be tensored with the pullback of $\cO_{\q}(k,l)$, giving the
decomposition 
\beq\label{deltak,l}
\Delta_* \obp(k,l) = \cO_{\q}(k,l) \oplus y \cO_{\q}(k-2,l-1)
\eeq
which will be the foundation for our cohomological calculations.

Let $S_x^k := H^0(\opx1(k))$ denote the $(k+1)$-dimensional
vector space of homogeneous polynomials of degree $k \ge 0$ in
$x_0,x_1$, with basis consisting of the monomials $x_0^k,
x_0^{k-1}x_1, \ldots, x_1^k$.
We set $S_x^k = 0$ for $k < 0$, and let $(S_x^k)^*$ denote the dual
vector space.
The cohomology of $\IP^1_x$ is given by:
\beq
H^0(\opx1(k)) = S_x^k, \quad
H^1(\opx1(k)) \simeq (S_x^{-2-k})^*,
\eeq
where the second formula involves Serre duality and therefore depends
on choosing, once and for all, an isomorphism of $K_{\IP^1_x}$ with
$\cO_{\IP^1_x}(-2)$.
This formula can be applied to the product surface $\q = \IP^1_t
\times \IP^1_x$, yielding a formula for the direct images
(for a general definition of direct image sheaves we refer the reader
to the Appendix)
\beq
R^i pr_{t*} \cO_{\q}(k,l) =
H^i(\opx1(k)) \otimes \opt1(l) \simeq
\left\{\ba{ll}
S_x^k \otimes \opt1(l), & i = 0 \\
(S_x^{-2-k})^* \otimes \opt1(l), & i = 1
\ea\right. \ ,
\eeq
and therefore for the cohomology:
\bea\label{Hn-qkl}
H^n(\cO_{\q}(k,l)) &=&
\bigoplus_{i+j=n} H^i(\opx1(k)) \otimes 
	H^j(\opt1(l))\nn\\
&\simeq&
\left\{\ba{ll}
S_x^k \otimes S_t^l, & n = 0 \\
(S_x^{-2-k})^* \otimes S_t^l \oplus S_x^k \otimes (S_t^{-2-l})^*, 
	& n = 1 \\
(S_x^{-2-k})^* \otimes (S_t^{-2-l})^* & n=2.
\ea\right.
\eea
The power of formula \eref{deltak,l} is that it allows us to write
down analogous formulas for the much more complicated surface $B'$:
\beq\label{bRb-ob}
\ba{ccccc}
\b'_* \obp(k,l) &=& S_x^k \otimes \opt1(l) &\oplus& 
	y S_x^{k-2} \otimes \opt1(l-1) \nn \\
R^1\b'_* \obp(k,l) &\simeq& (S_x^{-2-k})^* \otimes \opt1(l) &\oplus&
	y (S_x^{-k})^* \otimes \opt1(l-1).
\ea\eeq
Note that for $k>0$ only the $\b'_*$ term is non-zero, while for $k<0$
only the $R^1\b'_*$ term is non-zero.
The cohomology on $B'$ can be obtained from \eref{bRb-ob}, or directly
from \eref{deltak,l}:
\beq
H^n(\obp(k,l)) = H^n(\cO_{\q}(k,l)) \oplus 
	 y H^n(\cO_{\q}(k-2,l-1)),
\eeq
where the individual terms are given in \eref{Hn-qkl}.

Explicitly, this formula gives bases for the various cohomology groups
on $B'$ consisting of monomials in $t_0, t_1, x_0, x_1, y$.
For example:
\beq
\ba{ll}
H^0(\obp(0,1)): & t_0, t_1 \\
H^0(\obp(1,0)): & x_0, x_1 \\
H^0(\obp(3,0)): & x_0^3, x_0^2x_1, x_0x_1^2, x_1^3 \\
H^0(\obp(2,1)): & t_0x_0^2, t_0x_0x_1, t_0x_1^2,
		 t_1x_0^2, t_1x_0x_1, t_1x_1^2, y.
\ea
\eeq

%%%%%%%%%%%%%%%%%%%%%%%%%%%%%%%%%%%%%%%%%%%%%
% Computing the Cohomology
%%%%%%%%%%%%%%%%%%%%%%%%%%%%%%%%%%%%%%%%%%%%%%
Now, we are ready to calculate the cohomology groups which we need on
$\tx$.
%%%%%%%%%%%%%%%%%%% V_2
\paragraph{$\bullet$ \fbox{$V_2$}}
We have
\beq\label{b-w2=0}
\b_* \ol{W}_2 = \b_* W_2 = 0
\eeq 
since these sheaves are
torsion-free and vanish at a generic point.
We also have $R^1\b_* \ol{W}_2 = 0$
because it is supported on $\ol{C}_2 \cap e_9$, which is empty.
The long exact sequence induced from \eref{seqW2} therefore gives:
\beq
0 = \b_*\ol{W}_2 \to \b_* i_{f_\infty *}(G^*) \to R^1 \b_* W_2
	\to R^1 \b_* \ol{W}_2 \to 0,
\eeq
so $R^1 \b_* W_2 = \b_* i_{f_\infty *}(G^*)$.
The Leray spectral sequence for $\pi: \tx \to B'$ therefore gives:
\bea\label{H1V2-1}
	H^1(\tx, V_2) 
&=& H^1(\tx, \pi'^*W_2 \otimes \pi^*L_2)
= H^0(B', R^1\pi_*\pi'^*W_2 \otimes L_2) \nn \\
&=& H^0(B', \b'^*R^1\b_*W_2 \otimes L_2)
= H^0(f_\infty, G^*) \otimes H^0(f_0',L_2).
\eea
Note that $h^0(f_\infty,G^*) = 1$, $h^0(f_0', L_2) = 6$, hence
$h^1(\tx, V_2) = 6$.

%%%%%%%%%%%%%%%%%%% V_3
\paragraph{$\bullet$ \fbox{$V_3$}}
We again have $\b_*W_3=0$, so  for $i=0,1$:
\beq
H^i(\tx, V_3) =  H^0(B', \b'^*R^i\b_*W_3 \otimes L_3)
=  H^0(\IP^1, R^i\b_*W_3 \otimes \b_*' L_3) = 0,
\eeq
where we have used that $\b'_*L_3 = 0$, which holds since $L_3 \cdot
f' = -4 < 0$.

%%%%%%%%%%%%%%%%%%% V-tilde
\paragraph{$\bullet$ \fbox{$\tv$}}
The long exact sequence induced from \eref{v23} gives:
\beq\label{H1tv}
0 = H^0(\tx, V_3) \to H^1(\tx, V_2) \to H^1(\tx, \tv)
	\to H^1(\tx, V_3) = 0,
\eeq
so
$H^1(\tx, \tv) = H^1(\tx, V_2) = H^0(f_\infty, G^*) \otimes
H^0(f_0',L_2)$ by \eref{H1V2-1}.

%%%%%%%%%%%%%%%%%%% A^2 V_2
\paragraph{$\bullet$ \fbox{$\av_2$}}
From \eref{chernWi} we know that $\wedge^2W_2 = c_1(W_2) = -f$.
But $\pi'^*\cO_B(-f) \simeq \pi^* \obp(-f')$, since both pull back
from the same sheaf $\op1(-1)$ on $\IP^1$.
Therefore,
\beq
\wedge^2 V_2 = \pi'^* \wedge^2 W_2 \otimes \pi^* (L_2 ^{\otimes 2})
\simeq \pi^* \obp(6,-1).
\eeq
Combining this with:
\beq
\pi_* \cO_{\tx} = \obp, \quad
R^1 \pi_* \cO_{\tx} = \obp(-f')
\eeq
gives us formulas for the direct images of $\wedge^2 V_2$:
\beq
\pi_* \wedge^2 V_2 \simeq \obp(6,-1), \quad
R^1 \pi_* \wedge^2 V_2 \simeq \obp(6,-2).
\eeq
We then push on to $\IP^1$ as in \eref{bRb-ob}, and since $R^1 \b'_* =
0$ for $k=6$, we find:
\bea
(\b' \circ \pi)_* \av_2 &=& \b'_*(\pi_* \av_2) = \b'_* \obp(6,-1) 
	= S_x^6 \otimes \opt1(-1) \oplus y S_x^{4} \otimes \opt1(-2)
	\nn \\
R^1 (\b' \circ \pi)_* \av_2 &=& \b'_*(R^1 \pi_* \av_2) =
	\b'_* \obp(6,-2) = 
	S_x^6 \otimes \opt1(-2) \oplus y S_x^{4} \otimes \opt1(-3)
	\nn \\
R^2 (\b' \circ \pi)_* \av_2 &=& 0.
\eea
Since none of these sheaves have any global sections, we find the
cohomology on $\tx$ by taking $H^1$ of the images on $\IP^1_t$:
\beq\label{h1-av2}\ba{ll}
H^0(\tx, \av_2) = 0,	& h^0(\tx, \av_2) = 0, \\
H^1(\tx, \av_2) = y S_x^4,	& h^1(\tx, \av_2) = 5, \\
H^2(\tx, \av_2) = S_x^6 \oplus y S_x^4 \otimes (S_t^1)^*,	
	& h^2(\tx, \av_2) = 7+2\times 5 = 17, \\
H^3(\tx, \av_2) = 0,	& h^3(\tx, \av_2) = 0.
\ea\eeq

%%%%%%%%%%%%%%%%%%% A^2 V_2^*
\paragraph{$\bullet$ \fbox{$\av_2^*$}}
The cohomology of $\av_2^*$ can be obtained from that of $\av_2$ by
Serre duality.
Equivalently, we can apply the above procedure to $\av_2^* =
\pi^*\obp(-6,1)$, noting that for $k=-6$ all the $\b'_*$ terms 
in \eref{bRb-ob} vanish:
\beq
\pi_* \av_2^* = \obp(-6,1), \quad
R^1 \pi_* \av_2^* = \obp(-6,0).
\eeq
\bea
(\b' \circ \pi)_* \av_2^* &=& 0, \nn \\
R^1 (\b' \circ \pi)_* \av_2^* &=& R^1 \b'_*(\pi_* \av_2^*) =
	R^1 \b'_* \obp(-6,1) \nn \\
&=& 
	S_x^{4*} \otimes \opt1(1) \oplus y S_x^{6*} \otimes \opt1,
	\nn \\
R^2 (\b' \circ \pi)_* \av_2^* &=& R^1\b'_*(R^1\pi_* \av_2^*) 
	= R^1\b'_* \obp(-6,0) \nn \\ 
&=& S_x^{4*} \otimes \opt1 \oplus y S_x^{6*} \otimes \opt1(-1),
\eea
\beq\label{H-av2*}\ba{llll}
H^0(\tx, \av_2^*) &=& 0,	& h^0(\tx, \av_2^*) = 0, \\
H^1(\tx, \av_2^*) &=&
H^0(\IP^1_t, S_x^{4*} \otimes \opt1(1) \oplus 
	y S_x^{6*} \otimes \opt1) & \\
&=& S_x^{4*} \otimes S^1_t \oplus y S_x^{6*},
	& h^1(\tx, \av_2^*) = 5 \times 2 + 7 = 17, \\
H^2(\tx, \av_2^*) &=& 
H^0(\IP^1_t, S_x^{4*} \otimes \opt1 \oplus y S_x^{6*} \otimes
\opt1(-1)) & \\
&=&  S_x^{4*},	& h^2(\tx, \av_2^*) = 5, \\
H^3(\tx, \av_2^*) &=& 0,	& h^3(\tx, \av_2^*) = 0.
\ea\eeq

%%%%%%%%%%%%%%%%%%% V2 x V3^*
\paragraph{$\bullet$ \fbox{$\v23^*$}}
We recall that $C_2 = \ol{C}_2 \cup f_\infty$, and $W_2$ is related to
$\ol{W}_2$ by sequence \eref{seqW2}.
If we tensor \eref{seqW2} by $W_3^*$ and push to $\IP^1$ with $\b_*$,
we find
\beq\label{seqW2xW3*}
0 \to \b_*( i_{f_\infty *}G^* \otimes W_3^*) \to \cF \to \ol{\cF} \to
0, 
\eeq
where
\beq
\cF := R^1 \b_*(W_2 \otimes W_3^*), \quad
\ol{\cF} := R^1 \b_*(\ol{W}_2 \otimes W_3^*),
\eeq
and the last term in \eref{seqW2xW3*} is 0 because $G^*$ has degree $+1$
on $f_\infty$.
All the sheaves in \eref{seqW2xW3*} have finite support:
\begin{itemize}
\item[$-$] $\ol{\cF}$ is supported on $\b(\ol{C}_2 \cap C_3)$.
	If we choose things generically, $\ol{C}_2 \cap C_3$ will
consist of 12 points $p_j$ in $B'$, the image
$\b(\ol{C}_2 \cap C_3)$ will consist of 12 distinct points $\hat{p}_j
:= \b(p_j) \in \IP^1$, $j=1, \ldots, 12$, and $\ol{\cF}$ will decompose
as the sum of 12 rank 1 skyscraper sheaves $\cF_j$ near each
$\hat{p}_j$: $\ol{\cF} = \bigoplus_{j=1}^{12} \cF_j$.

\item[$-$] $\b_*(i_{f_\infty *}G^* \otimes W_3^*)$ is supported
	at $\infty \in \IP_t^1$, and has rank 3 there.
It can therefore be decomposed (non-canonically) as 
$\bigoplus_{j=13}^{15} \cF_j$, with each $\cF_j$ a rank 1 skyscraper
sheaf supported at $\infty$.
For $j=13,14,15$ we use $\hat{p}_j$ as another notation for the point
$\infty \in \IP_t^1$, the support of $\cF_j$.

\item[$-$] 
The sequence \eref{seqW2xW3*} splits, so 
$\cF = \bigoplus_{j=1}^{15} \cF_j$.

\end{itemize}

We can now combine this with formula \eref{bRb-ob} applied to $L_2
\otimes L_3^* = \obp(5,0)$, to compute $H^1(\tx, \v23^*)$:
\bea\label{h1v2v3*}
	H^1(\tx, \v23^*) 
&=& H^0(\IP^1_t, R^1\b_*(W_2\otimes W_3^*) \otimes \b_*'(L_2 \otimes
	L_3^*) ) \nn \\
&=& H^0(\IP^1_t, \cF \otimes [ S_x^5 \otimes \opt1 \oplus
	y S_x^3 \otimes \opt1(-1) ]) \nn \\
&=& \bigoplus_{j=1}^{15} H^0(\IP^1_t, \cF_j) \otimes
	[S_x^5 \oplus y S_x^3 \otimes \{\hat{p}_j\IC\} ].
\eea
Here, we use the notation $ \{\hat{p}_j\IC\} \subset S_t^{1*}$ for the
line inside the 2-dimensional plane  $S_t^{1*}$ consisting of all
points proportional to $\hat{p}_j \in \IP^1_t = \IP(S_t^{1*})$.
This line is the fiber at $\hat{p}_j$ of the line bundle $\opt1(-1)$. 
In particular, the dimension is
\beq
h^1(\tx, \v23^*) = 150 =  15 \times [6+4].
\eeq

%%%%%%%%%%%%%%%%%%% V2^* x V3^*
\paragraph{$\bullet$ \fbox{$V_2^* \otimes V_3^*$}}
From the Chern character formula \eref{chernWi} we know that $W_2^*
\simeq W_2 \otimes \obp(f)$, and therefore
\beq
R^1\b_*(W_2^*\otimes W_3^*) \simeq R^1\b_*(\b^* \opt1(1) \otimes W_2
	\otimes W_3^*)
= \cF \otimes \opt1(1).
\eeq
In analogy with \eref{h1v2v3*} we therefore get
\bea\label{H2-v2*v3*}
H^2(\tx, V_2^* \otimes V_3^*) &=&
H^0(\IP^1_t, R^1\b_*(W_2^*\otimes W_3^*) \otimes 
	R^1\b'_*(L_2^* \otimes L_3^*)) \nn \\
&=& H^0(\IP^1_t, \cF \otimes [y S_x^{1*}]) \nn \\
&=& \bigoplus_{j=1}^{15} 
	H^0(\IP^1_t, \cF_j) \otimes y S_x^{1*},
\eea
and the dimension is
\beq\label{h2v2*v3*}
h^2(\tx, V_2^* \otimes V_3^*) = 30 = 15 \times 2.
\eeq

%%----------------------------------
%\subsection{Computing $H^1(\tx, \avt)$}
%Having computed $H^1(\tilde{X}, U_i(\tilde{V}))$ for $U_i(\tilde{V}) =
%\tv$ and $\tv^*$, we proceed to the calculation for $U_i(\tilde{V}) =
%\avt$ and $\avt^*$.

\paragraph{$\bullet$ \fbox{$\avt$}}
We note that the short exact sequence \eref{v23}
which defines $\tv$ implies the exact sequence
\beq\label{defQ}
0 \to \av_2 \to \avt \to Q \to 0 \ ,
\eeq
where $Q$ is defined by the quotient of the map $\av_2 \to \avt$.
However, the natural map $\avt \to \av_3$ factors through $Q$ with the
kernel $V_2 \otimes V_3$. A simple consistency
check for this statement is by
dimension counting. Recall that $V_2$, $V_3$ and $\tv$ have rank 2, 3
and 5 respectively. Then,
$Q$ has dimension $\frac{5\cdot 4}{2} -
\frac{2\cdot 1}{2} = 9$ from \eref{defQ}, $\av_3$ has dimension
$\frac{3\cdot 2}{2} = 3$, so the kernel should have dimension
$9-3=6$. This is indeed the dimension of
$V_2 \otimes V_3$, which is $2 \cdot 3 = 6$.
In summary, we have an intertwined pair of short exact sequences as
follows.
\beq
\ba{cccccccccc}
&&&&&&0&&&\\
&&&&&&\uparrow&&&\\
&&&&&&\av_3&&&\\
&&&&&&\uparrow&&&\\
0 &\to& \av_2 &\to& \avt &\to& Q &\to& 0 \ . \\
&&&&&&\uparrow&&&\\
&&&&&&V_2 \otimes V_3&&&\\
&&&&&&\uparrow&&&\\
&&&&&&0&&&
\ea
\eeq

This then induces the following two long exact sequences in
cohomology,
\beq\label{seqavt1}
\ba{ccccccccc}
0 & \to & H^0(\tx, \av_2) & \to & H^0(\tx, \avt) & \to & H^0(\tx, Q) &
\to & \\
& \to & H^1(\tx, \av_2) & \to & \fbox{$H^1(\tx, \avt)$} & 
	\to & H^1(\tx, Q) & \to & \\
& \to & H^2(\tx, \av_2) & \to & H^2(\tx, \avt) & \to & H^2(\tx, Q) &
\to & \\
& \to & H^3(\tx, \av_2) & \to & H^3(\tx, \avt) & \to & H^3(\tx, Q) &
\to & 0 \ ,
\ea
\eeq
and
\beq\label{seqQ}
\ba{ccccccccc}
0 & \to & H^0(\tx, V_2 \otimes V_3) & \to & H^0(\tx, Q) & 
	\to & H^0(\tx, \av_3) & \to & \\
& \to & H^1(\tx, V_2 \otimes V_3) & \to & H^1(\tx, Q) & 
	\to & H^1(\tx, \av_3) & \to & \\
& \to & H^2(\tx, V_2 \otimes V_3) & \to & H^2(\tx, Q) & 
	\to & H^2(\tx, \av_3) & \to & \\
& \to & H^3(\tx, V_2 \otimes V_3) & \to & H^3(\tx, Q) & 
	\to & H^3(\tx, \av_3) & \to & 0 \ .
\ea
\eeq
We have boxed $H^1(\tx, \avt)$ since it is the term we wish
to compute.

First consider the second sequence \eref{seqQ}.
By the same arguments as \eref{b-w2=0}, we have that
\beq\label{H03v23}
H^0(\tx, V_2 \otimes V_3) = 
H^3(\tx, V_2 \otimes V_3) = 
H^0(\tx, \av_3) = 
H^3(\tx, \av_3) = 0 \ .
\eeq
It then follows from \eref{seqQ} that
\beq\label{h03Q0}
H^0(\tx, Q) = H^3(\tx, Q) = 0.
\eeq
Furthermore, using the Leray spectral sequence and the fact that 
$\pi_* \av_3 = 0$ implies
\beq\label{H1av3=H0}
H^1(\tx, \av_3) \simeq H^0(B', R^1\pi_* \av_3).
\eeq
Now,
\beq\label{R1piwV3}
R^1\pi_* \av_3 = \b^{'*}(R^1 \b_* \wedge^2 W_3) \otimes L_3^{\otimes
2}.
\eeq
Therefore, pushing \eref{R1piwV3} down to $\IP^1$, \eref{H1av3=H0}
becomes
\beq
H^0(B', R^1\pi_* \av_3) = H^0(\IP^1, (R^1 \b_* \wedge^2 W_3) \otimes
\b'_* L_3^{\otimes 2}).
\eeq
Using \eref{L23}, we see that $L_3^{\otimes 2}$ 
has negative degree along a generic fiber. Therefore, assuming that
the support of $R^1\b_* \wedge^2 W_3$ is on irreducible fibers,
$\b'_* L_3^{\otimes 2}$ vanishes and
\beq\label{H1av3}
H^1(\tx, \av_3) = 0 \ .
\eeq
Substituting \eref{H03v23} and \eref{H1av3} into \eref{seqQ} implies
\beq\label{H1Q}
H^1(\tx,Q) \simeq H^1(\tx, V_2\otimes V_3) \ ,
\eeq
and that $H^2(\tx,Q)$ fits into the short exact sequence
\beq
0 \to  H^2(\tx, V_2 \otimes V_3)  \to  H^2(\tx, Q)  
	\to  H^2(\tx, \av_3)  \to 0.
\eeq
Having established these results, let us now consider the first sequence
\eref{seqavt1}.
Substituting \eref{h03Q0} into \eref{seqavt1} gives
\beq
H^0(\tx, \avt) \simeq H^0(\tx, \av_2) \ ,
\eeq
and
\beq\label{seqavt2}
0 \to H^1(\tx,\av_2) \to \fbox{$H^1(\tx, \avt)$}  
\to  H^1(\tx, Q) \to H^2(\tx,\av_2) \to
\ldots
\eeq
Putting \eref{H1Q} into \eref{seqavt2} then leads to the exact
sequence
\beq\label{seqavt}
0 \to H^1(\tx,\av_2) \to 
\fbox{$H^1(\tx, \avt)$}  \to  H^1(\tx, V_2\otimes V_3)  
\stackrel{M^T}{\longrightarrow}
H^2(\tx,\av_2) \to \ldots
\eeq
with which we will determine the desired boxed term.
In \eref{seqavt}, we have explicitly labeled a map $M^T$,
namely the coboundary map 
\beq\label{defMt}
M^T :  H^1(\tx, \v23) \to H^2(\tx,\av_2) \ .
\eeq
It is given by cup product with
\beq\label{defMt2}
[\tv] \in H^1(\tx, V_3^* \otimes V_2) = 
\ext^1_{\tx}(V_3,V_2) \ ,
\eeq
the extension class of $\tv$, via the pairing
\beq
\ba{cccccc}
\cM^T: & H^1(\tx, \v23) &\times& 
H^1(\tx, V_3^* \otimes V_2)
&\to& H^2(\tx,\av_2) \\
& A & \times & B & \to & C \ .
\ea
\eeq
This can be dualized to
\beq\label{dualM}
\ba{cccccc}
\cM:&H^1(\tx,\av_2^*) &\times&
H^1(\tx, V_3^* \otimes V_2)
&\to&
H^2(\tx, V_2^* \otimes V_3^*) \\
& C^* & \times & B & \to & A^*
\ea \ .
\eeq

In formulas \eref{H-av2*}, \eref{h1v2v3*} and \eref{H2-v2*v3*} we have
expressed the three cohomology groups in \eref{dualM} as $H^0$ on
$\IP^1_t$ of appropriate sheaves.
The naturality of our construction implies that the multiplication map
$\cM$ on cohomologies is itself induced from the natural multiplication
map of the underlying sheaves on $\IP^1_t$, namely:
\beq
\left(S_x^{4*} \otimes \opt1(1) \oplus y S_x^{6*} \otimes \opt1\right)
\otimes
\left(\cF \otimes [ S_x^5 \otimes \opt1 \oplus
	y S_x^3 \otimes \opt1(-1)]\right)
\to
\cF \otimes y S_x^{1*}.
\eeq
By taking global sections, we find that $\cM$ is the product:
\beq\label{M-sec}
\cM: 
\left(S_x^{4*} \otimes S^1_t \oplus y S_x^{6*}\right) 
\otimes
\left(\bigoplus_{j=1}^{15} H^0(\IP^1_t, \cF_j) \otimes
	[S_x^5 \oplus y S_x^3 \otimes \{\hat{p}_j\IC\}]\right)
\to
\bigoplus_{j=1}^{15} H^0(\IP^1_t, \cF_j) \otimes y S_x^{1*}.
\eeq
In particular, our $\cM$ breaks into blocks.
The three spaces involved in $\cM$ have dimensions 17, 150 and 30
respectively.
This breaks into 15 blocks $(j=1,\ldots,15)$, each sending a $17
\times 10$ dimensional space to a 2-dimensional space.
Each block breaks further into a $10 \times 4 \to 2$ sub-block and a $7
\times 6 \to 2$ sub-block, corresponding to the products
\beq
(S_x^{4*} \otimes S_t^1) 
\otimes 
(S_x^{3} \otimes \{\hat{p}_j\IC\})
\to
S_x^{1*}
\eeq 
and
\beq
(S_x^{6*}) \otimes (S_x^{5}) \to S_x^{1*},
\eeq
respectively. (We have suppressed a $y H^0(\cF_j)$ factor on each side).
The transpose $M : C^* \to A^*$ of our map $M^T$ is obtained from 
\eref{M-sec} by evaluating at the extension class $[\tv] \in B$.
We can express this $[\tv]$ in terms of its coefficients
$a_{i,j}$, $i=0,\ldots,5$, $j=1,\ldots,15$ and $b_{k,j}$,
$k=0,\ldots,3$, $j=1,\ldots,15$, in the $S_x^5$ and $S_x^3$ factors
respectively.
Now the map $S_x^{6*} \to S_x^{1*}$ given by the $a_{i,j}$ is
represented by the $2 \times 6$ matrix
\beq
M_{I, j} = 
\left(
\ba{ccc}
a_{0,j} & \ldots~~a_{5,j} & 0 \\
0 & a_{0,j}~~\ldots & a_{5,j}
\ea
\right),
\eeq
while the map $S_x^{4*} \otimes S_t^1 \to S_x^{1*}$ given by the
$b_{k,j}$ is represented by the $2\times 10$ matrix
\beq
M_{II,j} = \left(\ba{c|c}
\ba{ccc}
b_{0,j}t_0(\hat{p}_j) & \ldots ~~ b_{3,j}t_0(\hat{p}_j)& 0 \\
0 & b_{0,j}t_0(\hat{p}_j) ~~\ldots & b_{3,j}t_0(\hat{p}_j)
\ea
&
\ba{ccc}
b_{0,j}t_1(\hat{p}_j) & \ldots ~~b_{3,j}t_1(\hat{p}_j) & 0 \\
0 & b_{0,j}t_1(\hat{p}_j) ~~\ldots & b_{3,j}t_1(\hat{p}_j)
\ea
\ea\right).
\eeq
So the full $30 \times 17$ matrix $M$ is then
\beq\label{M}
M = \left(\ba{ccc}
M_{I,1} & & M_{II,1} \\
\vdots  & & \vdots \\
M_{I,15} & & M_{II,15}
\ea\right).
\eeq
For a generic choice of the $a_{i,j}$ and $b_{k,j}$, the rank of $M$
is 17 and $M$ is surjective.
It is easy to see that this remains true also for generic
$\tau$-invariant extension $[\tv]$.
Plugging this, along with formulas \eref{H-av2*} and \eref{h2v2*v3*},
into \eref{seqavt}, we find:
\beq\label{avt=18}
h^1(\tx, \avt) = 5+30 - 17 = 18.
\eeq
Using Serre duality on $\tilde{X}$ and the fact that $\ind(\tilde 
V)=\ind(\wedge^2 \tv)=6$ \cite{spec}, 
it is now straightforward to determine the 
remaining cohomology groups of $\tilde{V}$, $\tilde{V}^*$, $\wedge^2
\tilde V$ and $\wedge^2 \tilde V^*$.

%%%%%
\section{The $\IZ_2$ Action}\label{s:z2}
In subsection \ref{s:tau} we described the involutions $\tau_B$,
$\tau_{B'}$, $\tau$ acting compatibly on $B$, $B'$ and $\tx$.
The action of $\tau_{B'}$ on line bundles on $B'$ is specified in
\tref{tab:tB}.
In particular, the line bundles $\obp(0,1)$ and $\obp(1,0)$ are
$\tau$-invariant.
It follows that there are induced involutions $\tau_{\IP^1_t}$,
$\tau_{\IP^1_x}$ that commute with the corresponding maps $\b':B' \to
\IP^1_t$, $\chi : B' \to \IP^1_x$.
We have already encountered the involution $\tau_{\IP^1_t}$ in
subsection \ref{s:B}, where we denoted it simply $\tau_{\IP^1}$.
It sends $t_0 \mapsto t_0, \quad t_1 \mapsto -t_1$.
We claim that $\tau_{\IP^1_x}$ is also a non-trivial involution, so
with an appropriate choice of the coordinates $x_0$, $x_1$ on
$\IP^1_x$ (note that we never fixed these coordinates up till now!) it
acts as $x_0 \mapsto x_0, \quad x_1 \mapsto -x_1$.
For this, we must determine the action of $\tau$ 
on the $\IP^1$ family of rational curves $r'$.
For a general, non-singular member of this family, all we learn from
\tref{tab:tB} is that it goes to another such.
But the table also tells us the image under $\tau_{B'}$ 
of each of the line
bundles $\obp(K_i^d)$, as $K_i^d$ runs over the 16 components of the 8
reducible curves in the system $|r'|$, specified in \eref{Kij}.
Each of these has the property that $K_i^d$ is the only effective
curve in its class: $h^0(B', K_i^d) = 1$.
So we can deduce from \tref{tab:tB} 
not only the cohomology class of the
image, but the actual physical image:
\beq
K_2^d \leftrightarrow K_3^d, \quad
K_1^d \leftrightarrow K_9^d, \quad
K_4^d \leftrightarrow K_7^{3-d}, \quad
K_6^d \leftrightarrow K_8^{3-d}.
\eeq
At any rate, this clearly demonstrates that $\tau_{\IP^1_x}$ is not
the identity, as claimed.

Via the map $\Delta$, our surface $B'$ is a double cover of $\q =
\IP^1_t \times \IP^1_x$.
Its equation can be written explicitly as
\beq
y^2 = F_{4,2}(x,t),
\eeq
with $F_{4,2}(x,t)$ a bihomogeneous polynomial, of degree 4 in
$x_0,x_1$ and of degree 2 in $t_0, t_1$.
By \eref{def-y}, $y$ is a section of $\obp(2,1)$ which vanishes on the
ramification locus $Ram_\Delta$.
Since $Ram_\Delta$ goes to itself under $\tau_{B'}$, $y$ must go to a
multiple of itself.
Since $\tau_{B'}$ is an involution, this multiple is $\pm 1$, so in
particular $F_{4,2}$ must be invariant (rather than anti-invariant).
From \eref{def-y}, it follows that either $\tau_{B'} y = y$ or 
$\iota \tau_{B'} y = y$.
Both involutions $\tau_{B'}$, $\iota \tau_{B'}$ have the same properties.
So by relabelling $\iota \tau_{B'}$ as $\tau_{B'}$ 
if necessary, we may as well
assume that the action of $\tau_{B'}$ is given explicitly by:
\beq\label{tx-trans}
t_0 \mapsto t_0, \quad t_1 \mapsto -t_1, \quad
x_0 \mapsto x_0, \quad x_1 \mapsto -x_1, \quad
y \mapsto y.
\eeq

In subsection \ref{s:V} we chose compatible actions of $\tau$ on
$V_2$, $V_3$ and $\tv$.
It turns out that the particle spectrum on $X$ is independent of these
choices and is precisely half the spectrum on $\tx$ which we computed
above. 
We compute it as follows.
%--------

\paragraph{$\bullet$ \fbox{$H^1(\tx, \tv)$}}
We have identified $H^1(\tx, \tv)$ with $H^0(f_\infty, G^*) \otimes
H^0(f_0',L_2)$ in \eref{H1V2-1}, \eref{H1tv}.
We plug $k=3$, $l=0$ into formula \eref{deltak,l}, and
restrict the double cover $\Delta : B' \to \q$ to $\chi: f_0' \to
\IP^1_x$, finding:
\beq
\chi_* \cO_{f_0'}(3r') = \opx1(3) \oplus y \opx1(1).
\eeq
We get a natural identification of $H^0(f_0', L_2) = H^0(f_0', 3r')$
with $S_x^3 \oplus y S_x^1$.
From \eref{tx-trans} we see that the $\tau$ action on this
6-dimensional space has a 3-dimensional invariant subspace and
3-dimensional anti-invariant subspace.
There is also a $\tau$-action on the 1-dimensional $H^0(f_\infty,
G^*)$, which must be either invariant or anti-invariant.
Either way, we find:
\beq\label{res-tv}
h^1(\tx, \tv)_+ = 3, \quad
h^1(\tx, \tv)_- = 3.
\eeq

%%%%%%%%%%%%%%%%%
\paragraph{$\bullet$ \fbox{$H^1(\tx, \avt)$}}
From the identification of $H^1(\tx, \av_2)$ with $y S_x^4$ in
\eref{h1-av2}, we see that
\beq
h^1(\tx, \av_2)_+ = 3, \quad
h^1(\tx, \av_2)_- = 2,
\eeq
while the identification of $H^2(\tx,\av_2)$ with 
$S_x^6 \oplus y S_x^4 \otimes (S_t^1)^*$ gives
\beq
h^2(\tx, \av_2)_+ = 4+5 = 9, \quad
h^2(\tx, \av_2)_- = 3+5 = 8.
\eeq
On the other hand, we saw in \eref{H2-v2*v3*} that $H^1(\tx, \v23)$ is
dual to $\bigoplus_{j=1}^{15} H^0(\IP^1_t, \cF_j) \otimes
(y S_x^{1*})$.
Again, the action of $\tau$ on the 2-dimensional space $y S_x^{1*}$
has 1-dimensional invariants and 1-dimensional anti-invariants, so
regardless of its action on the 15 1-dimensional spaces $H^0(\IP^1_t,
\cF_j)$, we get:
\beq
h^1(\tx, \v23)_+ = 15, \quad
h^1(\tx, \v23)_- = 15.
\eeq
Combining the last three formulae with \eref{seqavt} and recalling
that $M^T$ is $\tau$-equivariant (since it is cup product with the
class $[\tv]$, which was taken in subsection \ref{s:V} to be
$\tau$-invariant), we see that for those generic choices to which
\eref{avt=18} applies we have:
\beq\label{res-avt}
h^1(\tx, \avt)_+ = 3 + 15 - 9 = 9, \quad
h^1(\tx, \avt)_- = 2 + 15 - 8 = 9.
\eeq

%%%%%%%%%%%%%%%%
\paragraph{$\bullet$ \fbox{$H^1(\tx,\tv^*)$ and $H^1(\tx,\avt^*)$}}
The spectrum also requires the terms $H^1(\tilde{X}, \tv^*)$ and 
$H^1(\tilde{X}, \avt^*)$.
These can be obtained from the three-family condition (C3) in
\eref{cond}, in conjunction with the index theorem \eref{donagi4}, 
as well as Serre duality \eref{Serre} presented in the Appendix.
Together with the fact that $H^0(\tx, \tv)$, $H^0(\tx, \tv^*) =
H^3(\tx, \tv)^*$, $H^0(\tx, \avt)$, and $H^0(\tx, \avt^*) = H^3(\tx,
\avt)^*$ all vanish, we have that
\beq
-h^1(\tx, U_i(\tilde{V})) + h^1(\tx, U_i(\tilde{V^*})) 
= 6, \qquad
U_i(\tilde{V}) = \tv,~\avt \ .
\eeq
In fact, a $\IZ_2$-graded version of the index theorem implies the
stronger result that
\beq\label{z2index}
-h^1(\tx, U_i(\tilde{V}))_{(\pm)} + h^1(\tx, U_i(\tilde{V^*}))_{(\pm)} 
= 3, \qquad
U_i(\tilde{V}) = \tv,~\avt \ .
\eeq
Alternatively, we can think of it as
the index theorem applied to each of
the $\tau$-invariant and anti-invariant pieces of the cohomology.

Therefore, combining \eref{z2index} with \eref{res-tv}, we have that
\beq
h^1(\tx, \tv^*)_+ = 6, \quad
h^1(\tx, \tv^*)_- = 6.
\eeq
Similarly, combining \eref{z2index} with \eref{res-avt}, we have that
\beq
h^1(\tx, \avt)_+ = 12, \quad
h^1(\tx, \avt)_- = 12.
\eeq

Let us summarize the conclusions of the last two sections.
It is convenient to introduce the following notation.
Consider, for example, the cohomology group $H^1(\tx, \tv)$.
We showed in Section \ref{s:coh} and Section \ref{s:z2} that
$h^1(\tx,\tv)=6$ and $h^1(\tx,\tv)_{(+)}=h^1(\tx,\tv)_{(-)}=3$
respectively.
Henceforth, we will express both of these facts by writing
\beq
H^1(\tx,\tv) = \IC^3_{(+)} \oplus \IC^3_{(-)}.
\eeq
Using this notation, we encapsulate the results of Section \ref{s:coh}
and Section \ref{s:z2} in \tref{t:sum}.

\begin{table}[h]
%\[
%\ba{||c|c|c|c|c|c|c||}\hline\hline
%U_i & H^q(\tx, U_i(\tv)) & R_i & h^q(\tx, U_i(\tv))
%& A_j & \chi_{\rho_{H^q}(\IZ_2)} &  
%\IC_{(+)}^{r} \oplus \IC_{(-)}^s \\ \hline \hline
%
%24 & H^1(\tx,\ad \tv) & 1 & & & &
%	\\ \hline
%1 & H^0(\tilde{X}, \cO_{\tx}) & 24 & 1 & 0 & 0  & 
%	\IC_{(+)}^{1}
%	\\ \hline
%10 & H^1(\tilde{X}, \avt) & 5 & 18 & 0 & 0 & \IC_{(+)}^{9}
%	\\ \hline
%& & & & 1 & 1 & \IC_{(-)}^{9}
%	\\ \hline
%\ol{10} & H^1(\tilde{X}, \avt^*) & \ol 5 & 24 & 
%	0 & 0 & \IC_{(+)}^{12}
%	\\ \hline
%& & & & 1 & 1 & \IC_{(-)}^{12}
%	\\ \hline
%5 & H^1(\tilde{X}, \tv) & \ol{10} & 6 & 0 & 0 & \IC_{(+)}^{3}
%	\\ \hline
%& & & & 1 & 1 & \IC_{(-)}^{3}
%	\\ \hline
%\ol 5 & H^1(\tilde{X}, \tv^*) & 10 & 12 & 0 & 0 & \IC_{(+)}^{6}
%	\\ \hline
%& & & & 1 & 1 & \IC_{(-)}^{6}
%	\\ \hline  \hline
%\ea
%\]
\[
\ba{||c|c|c|c|c|c||}\hline\hline
U_i & H^q(\tx, U_i(\tv)) & R_i & h^q(\tx, U_i(\tv))
& A_j & \IC_{(+)}^{r} \oplus \IC_{(-)}^s \\ \hline \hline

1 & H^0(\tilde{X}, \cO_{\tx}) & 24 & 1 & 0   & 
	\IC_{(+)}^{1}
	\\ \hline
10 & H^1(\tilde{X}, \avt) & 5 & 18 & 0  & \IC_{(+)}^{9}
	\\ \hline
& & & & 1  & \IC_{(-)}^{9}
	\\ \hline
\ol{10} & H^1(\tilde{X}, \avt^*) & \ol 5 & 24 & 
	0 &  \IC_{(+)}^{12}
	\\ \hline
& & & & 1  & \IC_{(-)}^{12}
	\\ \hline
5 & H^1(\tilde{X}, \tv) & \ol{10} & 6 & 0 &  \IC_{(+)}^{3}
	\\ \hline
& & & & 1 &  \IC_{(-)}^{3}
	\\ \hline
\ol 5 & H^1(\tilde{X}, \tv^*) & 10 & 12 & 0 &  \IC_{(+)}^{6}
	\\ \hline
& & & & 1 &  \IC_{(-)}^{6}
	\\ \hline  \hline
\ea
\]
\caption{The dimensions and $\IZ_2$ actions on the cohomology spaces
$H^q(\tx, U_i(\tv))$.
%The $\chi_{\rho_{H^q}(\IZ_2)}$ are the characters of the $\IZ_2$
%representations.
}\label{t:sum}
\end{table}

%%%%---------
\section{Low Energy Spectrum}
We know from the discussion in Section 2, and specifically from
equation \eref{spec}, that the multiplicities of the representations 
$B_{ij}$ of the low energy
gauge group are determined by 
$(H^q(\tilde{X}, U_i(\tilde{V})) \otimes A_j)^{\rho'(F)}$, the
invariant part of $H^q(\tilde{X}, U_i(\tilde{V})) \otimes A_j$ under
the joint action of $\IZ_2$ on $H^q(\tilde{X}, U_i(\tilde{V}))$ and
$A_j$. 
By combining the results in \tref{tab:eg2} with the $\IZ_2$ action on
the cohomology groups listed in \tref{t:sum}, 
we can now compute the complete low energy spectrum of
our theory. 
The associated multiplets descend to $X = \tx / \IZ_2$ to form the
$\sm$ particle physics spectrum.
The results are listed in \tref{t:final}. The representation $R_i =
1$, corresponding to the moduli $H^0(\tx, \ad \tv)$, is not presented.

\begin{table}[h]
\[
\ba{||c|c|c|c||}\hline\hline
R_i & A_j &  
(H^q(\tilde{X}, U_i(\tilde{V})) \otimes A_j)^{\rho'(F)} & B_{ij}
\\ \hline \hline
%1 & & & \\ \hline
24 & 0  & \IC_{(+)}^{1} & 
	(8,1)_0 \oplus (1,3)_0 \oplus (1,1)_0
	\\ \hline
5 & 0 & \IC_{(+)}^{9} & (3,1)_{-2}
	\\ \hline
& 1 & \IC_{(-)}^{9} & (1,2)_{3}
	\\ \hline
\ol 5 & 0 & \IC_{(+)}^{12} & (\ol 3,1)_{2}
	\\ \hline
& 1 & \IC_{(-)}^{12} & (1, 2)_{-3}
	\\ \hline
\ol{10} & 0 & \IC_{(+)}^{3} & (3,1)_{4} \oplus (1,1)_{-6}
	\\ \hline
& 1 & \IC_{(-)}^{3} & (\ol 3, 2)_{-1}
	\\ \hline
10 & 0 & \IC_{(+)}^{6} & (\ol 3,1)_{-4} \oplus (1,1)_{6}
	\\ \hline
& 1 & \IC_{(-)}^{6} & (3, 2)_{1}
	\\ \hline 
\hline
\ea
\]
\caption{The particle spectrum of the low-energy $\sm$ theory.
The $A_j$ correspond to
characters of the $\IZ_2$ representations on $R_i$.
The $U(1)$ charges listed are $w=3Y$.
}
\label{t:final}
\end{table}

To begin with, the spectrum contains one copy of vector supermultiplets
transforming under $\sm$ as
\beq
(8,1)_0 \oplus  (1,3)_0 \oplus (1,1)_0.
\eeq
Contained in these multiplets are the gauge connections for $SU(3)_C$,
$SU(2)_L$ and $U(1)_Y$ respectively.
Furthermore, it contains three families of quarks and lepton
superfields, each family transforming as
\beq\label{ql}
(3,2)_{1}, \quad (\ol{3},1)_{-4}, \quad (\ol{3},1)_{2}
\eeq
and
\beq
(1,2)_{-3}, \quad (1,1)_{6}
\eeq
respectively.
Each of these multiplets is a chiral superfield, none of which has a
conjugate partner.
However, there are additional chiral superfields in the spectrum.
It follows from \tref{t:final} that these occur as conjugate pairs
%(the three chiral families is the difference, as dictated by the index
%theorem, between the cohomologies
%of $U(V)$ and $U(V^*)$, the remaining pair up) 
 of the $\sm$ representations
\beq\label{5}
(3,1)_{-2}, \quad (1,2)_{3}
\eeq
and
\beq\label{10bar}
(3,1)_{4} \oplus (1,1)_{-6}, \quad (\ol{3},2)_{-1}.
\eeq
These multiplets represent extra matter in the spectrum, such as Higgs
and other exotic fields.

Let us explain how the quark/lepton fermions and conjugate pairs arise.
Consider, for example, the $B_{ij}$ representations $(\ol{3},2)_{-1}$
and $(3,2)_{1}$, corresponding to the $\ol{10}$ and 10 representations
respectively. 
From \tref{t:final}, we see that there are 3 copies of
$(\ol{3},2)_{-1}$ and 6 copies of $(3,2)_{1}$.
Note that $6-3=3$ 
copies of $(3,2)_{1}$ are unpaired, as a consequence of
the index theorem.
Each unpaired $(3,2)_{1}$ multiplet contributes to a single
quark/lepton generation, as in \eref{ql}.
This leaves 3 conjugate pairs of $(\ol{3},2)_{-1}$ and $(3,2)_{1}$
superfields.
Being non-chiral pairs, these do not contribute to a quark/lepton
family but, rather, are additional supermultiplets
as listed in \eref{5} and \eref{10bar}.

%Similar remarks hold for all the other chiral supermultiplets.
%These multiplets arise as $\IZ_2$ invariants in the 5 and $\ol{10}$
%representations of $H = SU(5)$.
It remains to enumerate the number of additional superfields.
From \tref{t:final}, we see that the spectrum has 
\beq
n_{(3,1)_{-2}} = 9, \quad
n_{(1,2)_{3}} = 9
\eeq
and
\beq
n_{(3,1)_{4} \oplus (1,1)_{-6}} = 3, \quad
n_{(\ol 3,2)_{-1}} = 3
\eeq
copies of \eref{5} and \eref{10bar} respectively.
The multiplicity $n_{(1,2)_{3}}$ corresponds to the number of Higgs
doublet conjugate pairs in the low energy spectrum.
The remaining multiplets in \eref{5} and \eref{10bar} are exotic.

We conclude that the low energy spectrum of the simple,
representative model discussed in this paper includes the requisite
three chiral
families of quarks and leptons. Additionally, it naturally has
Higgs doublet supermultiplet pairs. 
Unfortunately, the spectrum contains extra, exotic
chiral supermultiplets which, potentially, are phenomenologically
unacceptable. 
%Whether they are or not depends to a large extent upon
%their Yukawa couplings and whether these coupling generate masses for
%these exotics at sufficiently high energy scale. 
However, these conjugate pairs of exotic multiplets may couple to the
moduli fields coming from $H^1(X, V \otimes V^*)$ to form mass terms. 
If the moduli can
acquire a sufficiently high vacuum expectation value, then the exotics
multiplets will decouple at low energy and be compatible with
phenomenology.
These couplings will be discussed
elsewhere. 

Armed with the technology developed in this paper, one can
now compute the spectra of standard-like models based on arbitrary
stable vector bundles on a wide range of elliptically fibered
Calabi-Yau threefolds. These spectra can be constrained to always
contain three families of quarks and leptons. We are presently
searching for such vacua with a phenomenologically acceptable number
of Higgs doublets and, hopefully, no exotic matter.

\paragraph{Acknowledgements}
We are grateful to Volker Braun and Tony Pantev for enlightening
discussions.
R.~D.~would like to acknowledge conversations with Jacques Distler.
This Research was supported in part by
the Department of Physics and the Maths/Physics Research Group
at the University of Pennsylvania
under cooperative research agreement DE-FG02-95ER40893
with the U.~S.~Department of Energy and an NSF Focused Research Grant
DMS0139799 for ``The Geometry of Superstrings.'' R.~D.~is partially
supported by an NSF grant DMS 0104354.
R.~R.~is also supported
by the Department of Physics and Astronomy of Rutgers University under
grant DOE-DE-FG02-96ER40959.

\appendix
\section{Some Useful Mathematical Facts}
In this Appendix, we present some useful mathematical facts used
throughout the paper \cite{Grif,hart,hart2}.
The first is Serre duality, which implies that for a sheaf $\cF$ on
an $n$-fold $X$
\beq\label{SerreGen}
H^q(X, \cF) \simeq H^{n-q}(X, \cF^* \otimes K_X)^*,
\eeq
where $K_X$ is the canonical bundle of $X$. For our Calabi-Yau
threefold $\tx$ and sheaf $U_i(\tilde{V})$ on $\tx$, \eref{SerreGen}
simplifies to
\beq\label{Serre}
H^q(\tx, U_i(\tilde{V})) \simeq H^{3-q}(\tx, U_i(\tilde{V})^*)^*,
\eeq
where we have used the fact that $K_{\tx}$ on 
a Calabi-Yau manifold is trivial.

The second tool we use is the 
Atiyah-Singer index theorem, which implies that on our Calabi-Yau
threefold $\tx$
\beq\label{index}
\ind(U_i(\tilde{V})
= \sum\limits_{q=0}^3 (-1)^q h^q(\tx, U_i(\tilde{V})) =
\int_{\tx} \ch(U_i(\tilde{V})) \wedge \td(\tx) = 
\frac12 \int_{\tx} c_3(U_i(\tilde{V})) \ .
\eeq

The three-generation condition means that on $X$, $\ind(V)$ is equal
to three \cite{GSW}, which implies that on the cover $\tx$
\cite{Donagi:2000zs,Donagi:2000zf},
\beq\label{3gen}
\ind(\tilde{V}) = |\IZ_2| \times 3 = 6,
\eeq
or,
\beq
c_3(\tilde{V}) = 12.
\eeq
This is the origin of the condition (C3) in \eref{cond}.

In this paper, we apply the index theorem in the two
cases $U_i(\tv) =\tv$ and $\avt$.
It was shown in Appendix A of \cite{spec} that for our $SU(5)$ bundle
$\tv$ 
\beq
c_3(\avt) = c_3(\tv) = 12.
\eeq
Therefore, in these cases, \eref{index} simplifies to
\beq\label{donagi4}
\sum\limits_{q=0}^3 (-1)^q h^q(\tx, U_i(\tilde{V})) 
%=\int_{\tx} \ch(U_i(\tilde{V})) \wedge \td(\tx) 
= 6, \qquad
U_i(\tilde{V}) = \tv,~\avt \ .
\eeq

An important tool for computing cohomology groups of vector
bundles or, more generally, coherent sheaves on fibered
spaces is the Leray spectral sequence. Consider the map
$\pi: \tx \to B'$ and a sheaf $\cF$ on $\tx$. The Leray spectral
sequence for the map $\pi$ will relate the cohomologies of $\cF$ on
$\tx$ to the cohomologies of the higher direct
image sheaves $R^i \pi_*\cF$ on
$B'$. For a general map, the sequence is rather complicated. However,
in the case of $\pi$ being an elliptic fibration, it will degenerate
to a simpler long exact sequence.

To begin with, consider the definition of $R^0\pi_*\cF = \pi_* \cF$. It
is a sheaf on $B'$ given by
\beq\label{piF}
\pi_* \cF (U) = \cF( \pi^{-1} (U)) = H^0(\pi^{-1}(U), \cF|_{\pi^{-1}(U)})
\eeq
for any open set $U \subset B'$. 
%In particular, if we consider the
%limit where $U$ degenerates to a point $b$, we find
%\beq\label{piFb}
%\pi_* \cF|_b = H^0(\pi^{-1}(b), \cF|_{\pi^{-1}(b)}) = H^0(f, \cF|_f) \
%, 
%\eeq
%where $f = \pi^{-1}(b)$.
The definition \eref{piF} generalizes to the higher image sheaves as
\beq\label{RipiFU}
R^i \pi_* \cF(U) = H^i(\pi^{-1}(U), \cF|_{\pi^{-1}(U)}) \ ,
\eeq
for sufficiently small $U$.
It follows that for the map $\pi: \tx \to B'$
\beq\label{RipiF}
R^i \pi_* \cF(U) = 0, \qquad i > \dim \pi^{-1}(U) \ .
\eeq
%Taking the limit of $U$ to a point $b$, \eref{RipiFU} becomes
%\beq
%R^i \pi_* \cF|_b = H^i(f, \cF|_f) = 0, \qquad i > 1 \ ,
%\eeq
%where we have used \eref{piFb} and the fact that 
%the fibers $f = \pi^{-1}(b)$ are tori and thus of dimension one. 
%It follows from \eref{RipiF} that the
In our case, the Leray spectral sequence degenerates to the long exact
sequence 
\beq\label{leray}
\ba{ccccccccc}
0 & \to & H^1(B', \pi_*\cF) & \to & H^1(\tx, \cF) & \to & 
	H^0(B',R^1\pi_*\cF)  & \to & \\
& \to & H^2(B', \pi_*\cF) & \to & H^2(\tx, \cF) & \to & 
	H^1(B',R^1\pi_*\cF) & \to & 0 \ .
%& \to & H^3(B', \pi_*\cF) & \to & H^3(\tx, \cF) & \to & 
%	H^2(B',R^1\pi_*\cF) & \to & 0 \ .
\ea
\eeq
Note that $H^3(B', \pi_* \cF) = 0$ since $\dim_\IC B' = 2$. As
promised, \eref{leray} relates the cohomology of $\cF$ on $\tx$ to the
cohomology of the higher image sheaves $R^i\pi_* \cF$ on $B'$. Recall
that $B'$ is itself elliptically fibered. Therefore, one can
write a Leray spectral sequence for the map $\beta' : B' \to \IP^1$ in
complete analogy to \eref{leray}.

Another useful formula is \GRR~(GRR),  which states
that for any map $f : X \rightarrow B$
and any sheaf $\cS$ on $X$, we have
\beq
%\label{thmGRR}
\td (TB) \mbox{ch}(\sum_{i=0}^2 (-1)^i R^if_* \cS)
=
f_*(\mbox{ch}(\cS) \td(TX)) \ .
\eeq

%================
%================
\bibliographystyle{JHEP}

\begin{thebibliography}{99}
\bibitem{GSW} M.~Green, J.~Schwarz and E.~Witten,
``Superstring theory, vol I \& II,''
Cambridge University Press, 1988

\bibitem{FMW1}
   R.~Friedman, J.~Morgan and E.~Witten,
   ``Vector bundles and F theory,''
   Commun.\ Math.\ Phys.\  {\bf 187}, 679 (1997).
   [hep-th/9701162].

\bibitem{D} R.~Donagi, 
``Principal bundles on elliptic fibrations,''
{\em Asian J. Math.}, 1(2):214--223, 1997, alg-geom/9702002.

\bibitem{FMW2}
R.~Friedman, J.~Morgan, and E.~Witten,
``Vector bundles over elliptic fibrations,''
{\em J. Algebraic Geom.}, 8(2):279--401, 1999,
alg-geom/9709029.

\bibitem{Donagi:1998xe}
R.~Donagi, A.~Lukas, B.~A. Ovrut, and D.~Waldram, {\it Non-perturbative vacua
  and particle physics in {M}-theory},  {\em JHEP} {\bf 05} (1999) 018,
  [\href{http://xxx.lanl.gov/abs/hep-th/9811168}{{\tt hep-th/9811168}}].

\bibitem{Lukas:1997fg}
A.~Lukas, B.~A. Ovrut, and D.~Waldram, {\it On the four-dimensional effective
  action of strongly coupled heterotic string theory},  {\em Nucl. Phys.} {\bf
  B532} (1998) 43--82, [\href{http://xxx.lanl.gov/abs/hep-th/9710208}{{\tt
  hep-th/9710208}}].

\bibitem{Lukas:1998ew}
A.~Lukas, B.~A. Ovrut, and D.~Waldram, {\it The ten-dimensional effective
  action of strongly coupled heterotic string theory},  {\em Nucl. Phys.} {\bf
  B540} (1999) 230--246, [\href{http://xxx.lanl.gov/abs/hep-th/9801087}{{\tt
  hep-th/9801087}}].

\bibitem{Lukas:1998hk}
A.~Lukas, B.~A. Ovrut, and D.~Waldram, {\it Non-standard embedding and
  five-branes in heterotic {M}-theory},  {\em Phys. Rev.} {\bf D59} (1999)
  106005, [\href{http://xxx.lanl.gov/abs/hep-th/9808101}{{\tt
  hep-th/9808101}}].

\bibitem{Lukas:1998tt}
A.~Lukas, B.~A. Ovrut, K.~S. Stelle, and D.~Waldram, {\it Heterotic {M}-theory
  in five dimensions},  {\em Nucl. Phys.} {\bf B552} (1999) 246--290,
  [\href{http://xxx.lanl.gov/abs/hep-th/9806051}{{\tt hep-th/9806051}}].

\bibitem{Andreas:1999ty}
B.~Andreas, G.~Curio and A.~Klemm,
``Towards the standard model spectrum from elliptic Calabi-Yau,''
Int.\ J.\ Mod.\ Phys.\ A {\bf 19}, 1987 (2004)
[arXiv:hep-th/9903052].


\bibitem{Diaconescu:1998kg}
D.~E.~Diaconescu and G.~Ionesei,
``Spectral covers, charged matter and bundle cohomology,''
JHEP {\bf 9812}, 001 (1998)
[arXiv:hep-th/9811129].


\bibitem{Lukas:1998yy}
A.~Lukas, B.~A. Ovrut, K.~S. Stelle, and D.~Waldram, {\it The universe as a
  domain wall},  {\em Phys. Rev.} {\bf D59} (1999) 086001,
  [\href{http://xxx.lanl.gov/abs/hep-th/9803235}{{\tt hep-th/9803235}}].

\bibitem{Donagi:1999gc}
R.~Donagi, A.~Lukas, B.~A. Ovrut, and D.~Waldram, {\it Holomorphic vector
  bundles and non-perturbative vacua in {M}- theory},  {\em JHEP} {\bf 06}
  (1999) 034, [\href{http://xxx.lanl.gov/abs/hep-th/9901009}{{\tt
  hep-th/9901009}}].

\bibitem{Lukas:1999kt}
A.~Lukas, B.~A. Ovrut, and D.~Waldram, {\it Five-branes and supersymmetry
  breaking in {M}-theory},  {\em JHEP} {\bf 04} (1999) 009,
  [\href{http://xxx.lanl.gov/abs/hep-th/9901017}{{\tt hep-th/9901017}}].

\bibitem{Lalak1}
Z. Lalak, S. Pokorski and S. Thomas,
{\em Beyond the Standard Embedding in M-Theory on $S^1/Z_2$},
Nucl.Phys. B549 (1999) 63-97 [hep-ph/9807503].

\bibitem{Buchbinder:2002pr}
E.~I. Buchbinder, R.~Donagi, and B.~A. Ovrut, {\it Vector bundle moduli
  superpotentials in heterotic superstrings and {M}-theory},  {\em JHEP} {\bf
  07} (2002) 066, [\href{http://xxx.lanl.gov/abs/hep-th/0206203}{{\tt
  hep-th/0206203}}].

\bibitem{Buchbinder:2002ji}
E.~Buchbinder, R.~Donagi, and B.~A. Ovrut, {\it Vector bundle moduli and small
  instanton transitions},  {\em JHEP} {\bf 06} (2002) 054,
  [\href{http://xxx.lanl.gov/abs/hep-th/0202084}{{\tt hep-th/0202084}}].

\bibitem{moduli}
Y.~H.~He, B.~A.~Ovrut and R.~Reinbacher,
``The moduli of reducible vector bundles,''
JHEP {\bf 0403}, 043 (2004)
[arXiv:hep-th/0306121].

\bibitem{moduli2}
E.~I.~Buchbinder, B.~A.~Ovrut and R.~Reinbacher,
``Instanton moduli in string theory,''
arXiv:hep-th/0410200.


\bibitem{Donagi:1999jp}
R.~Donagi, B.~A. Ovrut, and D.~Waldram, {\it Moduli spaces of fivebranes on
  elliptic Calabi-Yau threefolds},  {\em JHEP} {\bf 11} (1999) 030,
  [\href{http://xxx.lanl.gov/abs/hep-th/9904054}{{\tt hep-th/9904054}}].



\bibitem{Ovrut:2000qi}
B.~A. Ovrut, T.~Pantev, and J.~Park, {\it Small instanton transitions in
  heterotic {M}-theory},  {\em JHEP} {\bf 05} (2000) 045,
  [\href{http://xxx.lanl.gov/abs/hep-th/0001133}{{\tt hep-th/0001133}}].

\bibitem{Kachru}
S. Kachru and E. Silverstein,
{\em Chirality Changing Phase Transitions in 4d String Vacua},
Nucl.Phys. B504 (1997) 272-284 [hep-th/9704185].

\bibitem{Curio}
G. Curio, {\em Chiral matter and transitions in heterotic string models},
Phys.Lett. B435 (1998) 39-48 [hep-th/9803224].


\bibitem{Douglas:2004yv}
M.~R.~Douglas and C.~g.~Zhou,
``Chirality change in string theory,''
JHEP {\bf 0406}, 014 (2004) [arXiv:hep-th/0403018].




\bibitem{Buchbinder:2002ic}
E.~I. Buchbinder, R.~Donagi, and B.~A. Ovrut, {\it Superpotentials for vector
  bundle moduli},  {\em Nucl. Phys.} {\bf B653} (2003) 400--420,
  [\href{http://xxx.lanl.gov/abs/hep-th/0205190}{{\tt hep-th/0205190}}].

\bibitem{Lima:2001nh}
E.~Lima, B.~A. Ovrut, and J.~Park, {\it Five-brane superpotentials in heterotic
  {M}-theory},  {\em Nucl. Phys.} {\bf B626} (2002) 113--164,
  [\href{http://xxx.lanl.gov/abs/hep-th/0102046}{{\tt hep-th/0102046}}].

\bibitem{Lima:2001jc}
E.~Lima, B.~A. Ovrut, J.~Park, and R.~Reinbacher, {\it Non-perturbative
  superpotential from membrane instantons in heterotic {M}-theory},  {\em Nucl.
  Phys.} {\bf B614} (2001) 117--170,
  [\href{http://xxx.lanl.gov/abs/hep-th/0101049}{{\tt hep-th/0101049}}].


\bibitem{Krause:2000gp}
A.~Krause, {\it A small cosmological constant, grand unification and warped
  geometry},  \href{http://xxx.lanl.gov/abs/hep-th/0006226}{{\tt
  hep-th/0006226}}.

\bibitem{Curio:2000dw}
G.~Curio and A.~Krause, {\it Four-flux and warped heterotic {M}-theory
  compactifications},  {\em Nucl. Phys.} {\bf B602} (2001) 172--200,
  [\href{http://xxx.lanl.gov/abs/hep-th/0012152}{{\tt hep-th/0012152}}].

\bibitem{Curio:2001ae}
G.~Curio, A.~Klemm, B.~Kors and D.~Lust,
``Fluxes in heterotic and type II string compactifications,''
Nucl.\ Phys.\ B {\bf 620}, 237 (2002)
[arXiv:hep-th/0106155].

\bibitem{Curio:2001qi}
G.~Curio and A.~Krause, {\it G-fluxes and non-perturbative stabilisation of
  heterotic {M}-theory},  {\em Nucl. Phys.} {\bf B643} (2002) 131--156,
  [\href{http://xxx.lanl.gov/abs/hep-th/0108220}{{\tt hep-th/0108220}}].

\bibitem{Curio:2003ur}
G.~Curio and A.~Krause, {\it Enlarging the parameter space of heterotic
  {M}-theory flux compactifications to phenomenological viability},  {\em Nucl.
  Phys.} {\bf B693} (2004) 195--222,
  [\href{http://xxx.lanl.gov/abs/hep-th/0308202}{{\tt hep-th/0308202}}].

\bibitem{Cardoso:2003af}
G.~L. Cardoso, G.~Curio, G.~Dall'Agata, and D.~L{\"u}st, {\it {BPS} action and
  superpotential for heterotic string compactifications with fluxes},  {\em
  JHEP} {\bf 10} (2003) 004,
  [\href{http://xxx.lanl.gov/abs/hep-th/0306088}{{\tt hep-th/0306088}}].

\bibitem{Cardoso:2002hd}
G.~L. Cardoso, G.~Curio, G.~Dall'Agata, P.~M. D.~L{\"u}st, and G.~Zoupanos,
  {\it Non-{K\"ahler} string backgrounds and their five torsion classes},  {\em
  Nucl. Phys.} {\bf B652} (2003) 5--34,
  [\href{http://xxx.lanl.gov/abs/hep-th/0211118}{{\tt hep-th/0211118}}].

\bibitem{spec1} 
R.~Donagi, Y.~H.~He, B.~A.~Ovrut and R.~Reinbacher,
``Moduli dependent spectra of heterotic compactifications,''
arXiv:hep-th/0403291.

\bibitem{spec}
R.~Donagi, Y.~H.~He, B.~A.~Ovrut and R.~Reinbacher,
``The particle spectrum of heterotic compactifications,''
arXiv:hep-th/0405014.




\bibitem{Khoury:2001zk}
J.~Khoury, B.~A. Ovrut, P.~J. Steinhardt, and N.~Turok, {\it Density
  perturbations in the ekpyrotic scenario},  {\em Phys. Rev.} {\bf D66} (2002)
  046005, [\href{http://xxx.lanl.gov/abs/hep-th/0109050}{{\tt
  hep-th/0109050}}].

\bibitem{Khoury:2001bz}
J.~Khoury, B.~A. Ovrut, N.~Seiberg, P.~J. Steinhardt, and N.~Turok, {\it From
  big crunch to big bang},  {\em Phys. Rev.} {\bf D65} (2002) 086007,
  [\href{http://xxx.lanl.gov/abs/hep-th/0108187}{{\tt hep-th/0108187}}].

\bibitem{Donagi:2001fs}
R.~Y. Donagi, J.~Khoury, B.~A. Ovrut, P.~J. Steinhardt, and N.~Turok, {\it
  Visible branes with negative tension in heterotic {M}-theory},  {\em JHEP}
  {\bf 11} (2001) 041, [\href{http://xxx.lanl.gov/abs/hep-th/0105199}{{\tt
  hep-th/0105199}}].

\bibitem{Khoury:2001wf}
J.~Khoury, B.~A. Ovrut, P.~J. Steinhardt, and N.~Turok, {\it The ekpyrotic
  universe: Colliding branes and the origin of the hot big bang},  {\em Phys.
  Rev.} {\bf D64} (2001) 123522,
  [\href{http://xxx.lanl.gov/abs/hep-th/0103239}{{\tt hep-th/0103239}}].

\bibitem{Sen:1985eb}
A.~Sen, ``The Heterotic String In Arbitrary Background Field,''
Phys.\ Rev.\ D {\bf 32}, 2102 (1985).

\bibitem{Witten:1985xc}
E.~Witten, ``Symmetry Breaking Patterns In Superstring Models,''
Nucl.\ Phys.\ B {\bf 258}, 75 (1985).

\bibitem{Breit:1985ud}
J.~D.~Breit, B.~A.~Ovrut and G.~C.~Segre,
``E(6) Symmetry Breaking In The Superstring Theory,''
Phys.\ Lett.\ B {\bf 158}, 33 (1985).

\bibitem{Evans:1985vb}
M.~Evans and B.~A.~Ovrut,
``Splitting The Superstring Vacuum Degeneracy,''
Phys.\ Lett.\ B {\bf 174}, 63 (1986).

\bibitem{Greene:1986ar}
B.~R.~Greene, K.~H.~Kirklin, P.~J.~Miron and G.~G.~Ross,
``A Superstring Inspired Standard Model,''
Phys.\ Lett.\ B {\bf 180}, 69 (1986).

\bibitem{Ruip}
B. Andreas and D. Hernandez Ruiperez,
{\em Comments on N=1 Heterotic String Vacua},
Adv.Theor.Math.Phys. 7 (2004) 751-786 [hep-th/0305123].

\bibitem{Andreas}
B. Andreas, {\em On Vector Bundles and Chiral Matter in N=1 Heterotic
Compactifications},
JHEP 9901 (1999) 011 [hep-th/9802202].

%%%%%%%%%%%DOPW PAPERS

\bibitem{Donagi:2000fw}
R.~Donagi, B.~A. Ovrut, T.~Pantev, and D.~Waldram, {\it Spectral involutions on
  rational elliptic surfaces},  {\em Adv. Theor. Math. Phys.} {\bf 5} (2002)
  499--561, [\href{http://xxx.lanl.gov/abs/math.ag/0008011}{{\tt
  math.ag/0008011}}].

\bibitem{Donagi:2000zs}
R.~Donagi, B.~A. Ovrut, T.~Pantev, and D.~Waldram, {\it Standard-model
  bundles},  {\em Adv. Theor. Math. Phys.} {\bf 5} (2002) 563--615,
  [\href{http://xxx.lanl.gov/abs/math.ag/0008010}{{\tt math.ag/0008010}}].

\bibitem{Donagi:2000zf}
R.~Donagi, B.~A. Ovrut, T.~Pantev, and D.~Waldram, {\it Standard-model bundles
  on non-simply connected Calabi-Yau threefolds},  
{\em JHEP} {\bf 08} (2001) 053,
  [\href{http://xxx.lanl.gov/abs/hep-th/0008008}{{\tt hep-th/0008008}}].

\bibitem{Donagi:1999ez}
R.~Donagi, B.~A. Ovrut, T.~Pantev, and D.~Waldram, {\it Standard models from
  heterotic {M}-theory},  {\em Adv. Theor. Math. Phys.} {\bf 5} (2002) 93--137,
  [\href{http://xxx.lanl.gov/abs/hep-th/9912208}{{\tt hep-th/9912208}}].


\bibitem{Donagi:2003tb}
R.~Donagi, B.~A. Ovrut, T.~Pantev, and R.~Reinbacher, {\it {$SU(4)$} instantons
  on calabi-yau threefolds with {$\IZ_2\times\IZ_2$} fundamental group},  {\em
  JHEP} {\bf 01} (2004) 022,
  [\href{http://xxx.lanl.gov/abs/hep-th/0307273}{{\tt hep-th/0307273}}].

\bibitem{Ovrut:2003zj}
B.~A. Ovrut, T.~Pantev, and R.~Reinbacher, {\it Invariant homology on standard
  model manifolds},  {\em JHEP} {\bf 01} (2004) 059,
  [\href{http://xxx.lanl.gov/abs/hep-th/0303020}{{\tt hep-th/0303020}}].

\bibitem{Ovrut:2002jk}
B.~A. Ovrut, T.~Pantev, and R.~Reinbacher, {\it Torus-fibered
Calabi-Yau threefolds
  with non-trivial fundamental group},  {\em JHEP} {\bf 05} (2003) 040,
  [\href{http://xxx.lanl.gov/abs/hep-th/0212221}{{\tt
hep-th/0212221}}].

\bibitem{Braun:2004xv}
V.~Braun, B.~A.~Ovrut, T.~Pantev and R.~Reinbacher,
``Elliptic Calabi-Yau threefolds with Z(3) x Z(3) Wilson lines,''
arXiv:hep-th/0410055.


\bibitem{GS} M.~B.~Green and J.~H.~Schwarz,
``Anomaly Cancellation In Supersymmetric D=10 Gauge Theory And
Superstring Theory,''
Phys.\ Lett.\ B {\bf 149}, 117 (1984).


\bibitem{schoen}
C.~Schoen,
``On fiber products of rational elliptic surfaces with section,''
{\em Math. Z.}, 197(2):177--199, 1988.


%\bibitem{mike} M.~Douglas, private communications.

\bibitem{Ev:moduli}
E.~Buchbinder, R.~Donagi and B.~A.~Ovrut,
``Vector bundle moduli and small instanton transitions,''
JHEP {\bf 0206}, 054 (2002)
[arXiv:hep-th/0202084].


\bibitem{letter}
R.~Donagi, Y.~H.~He, B.~Ovrut and R.~Reinbacher,
``Higgs doublets, split multiplets and heterotic $SU(3)_C x SU(2)_L x
U(1)_Y$ spectra,''
arXiv:hep-th/0409291.

\bibitem{Grif}
P. Griffiths and J. Harris,
{\em Principles of Algebraic Geometry},
John Wiley and Sons, 1994.

\bibitem{hart}
R. Hartshorne, ``Algebraic Geometry,'' Springer-Verlag, 1997.

\bibitem{hart2}
Robin Hartshorne, ``Residues and Duality,''
Springer-Verlag, New York, 1966 

\end{thebibliography}

\end{document}